\documentclass[fleqn,10pt]{wlscirep}
\usepackage[utf8]{inputenc}
\usepackage[T1]{fontenc}
\usepackage{graphicx} 
\usepackage{filecontents}
\usepackage{multirow}

\title{Staining normalization in histopathology: Method benchmarking using multicenter dataset}

\author[1]{Umair Khan}
\author[2]{Jouni Härkönen}
\author[2]{Marjukka Friman}
\author[3,1]{Leena Latonen}
\author[2,]{Teijo Kuopio}
\author[1,4,5,*]{Pekka Ruusuvuori}
\affil[1]{University of Turku, Institute of Biomedicine, Turku, FI-20014, Finland}
\affil[2]{Department of Pathology, Hospital Nova of Central Finland, Jyväskylä, Finland}
\affil[3]{University of Eastern Finland, Institute of Biomedicine, Kuopio, Finland}
\affil[4]{Tampere University, Faculty of Medicine and Health Technology, Tampere, 33100, Finland }
\affil[5]{InFlames Research Flagship, University of Turku, Turku, Finland}

\affil[*]{corresponding author(s): Pekka Ruusuvuori (pekka.ruusuvuori@utu.fi)}

\begin{abstract}
Hematoxylin and Eosin (H\&E) has been the gold standard in tissue analysis for decades, however, tissue specimens stained in different laboratories vary, often significantly, in appearance. This variation poses a challenge for both pathologists' and AI-based downstream analysis. Minimizing stain variation computationally is an active area of research. To further investigate this problem, we collected a unique multi-center tissue image dataset, wherein tissue samples from colon, kidney, and skin tissue blocks were distributed to 66 different labs for routine H\&E staining. To isolate staining variation, other factors affecting the tissue appearance were kept constant. Further, we used this tissue image dataset to compare the performance of eight different stain normalization methods, including four traditional methods, namely, histogram matching, Macenko, Vahadane, and Reinhard normalization, and two deep learning-based methods namely CycleGAN and Pixp2pix, both with two variants each. We used both quantitative and qualitative evaluation to assess the performance of these methods. The dataset's inter-laboratory staining variation could also guide strategies to improve model generalizability through varied training data.
\end{abstract}
\begin{document}

\flushbottom
\maketitle

\section*{Introduction}
Tissue slide digitization has helped histological analysis tremendously over the decades. It has expedited the whole process by reducing the need of manual slide handling during observation and also allowed us to archive tissue specimens in a more efficient manner. Once the tissue specimen is scanned, the digital image is stored and kept safe from any physical changes or deterioration. Tissue processing and staining are complex processes that have a significant effect on tissue appearance. Typically, tissue specimens go through fixation, embedding, and sectioning before staining. Factors such as over- or under-fixation, choice of embedding medium and temperature, and thickness of the tissue section contribute to the varying appearance of tissue images \cite{troiano2009effects, chlipala2021impact}. Once processed, the tissue sections are then stained. For instance, one of the most commonly used stain combinations is Hematoxylin and Eosin (H\&E) that reveals the intricate morphology of the tissue specimen which is otherwise indiscernible prior to staining. Hematoxylin imbues the cell nuclei with shades of a purple-blue color and Eosin stains cytoplasm and extracellular matrix with shades of a pink-red color \cite{chan2014wonderful}. Even though H\&E is the most commonly used stain combination in routine diagnostics, differences in stain manufacturer, composition and staining protocol manifest in varying appearance of the stained tissue across different laboratories \cite{bancroft2008theory}. The appearance of the stained tissue is prone to further variation during the imaging phase, as differences in microscope/scanner hardware, image acquisition and post processing techniques from different manufacturers add to the variation \cite{weinstein2009overview, ji2023physical}. The combined effect of these factors pose challenges for post digitalization tasks. For example, we have previously shown that differences in appearance due to fixatives contribute towards differences in performance of image analysis, specifically, in nucleus detection \cite{valkonen2020generalized}. 

Digitization of histopathological tissue samples into high resolution whole slide images has enabled the development and use of computer assisted tools based on artificial intelligence (AI) for diagnostics and decision support \cite{rakha2021current, moscalu2023histopathological}. These diagnostic tools have the potential to improve efficiency in healthcare as well as provide access to diagnostic services for regions that have been suffering from a chronic shortage of pathologists \cite{wilson2018access}. Studies have shown that use of AI in digital pathology can be highly effective and accurate on tasks such as cancer grading, tumor cell segmentation, prognostication, and biomarkers prediction \cite{strom2020artificial, balkenhol2019deep, zadeh2021deep, vu2019methods, wulczyn2020deep, saillard2020predicting, fu2020pan, ahmed2022deep}, but gaining regulatory approval for such digital tools is a long-drawn and arduous journey. Only a handful of these efforts have successfully translated into AI-assisted diagnostic tools approved by regulatory authorities for clinical use \cite{paigeai}. One of the biggest challenges in developing tools that can be widely applied is their generalization capability, which is often significantly hindered by the high variation of tissue appearance stemming from the sample preparation, staining, and imaging \cite{dunn2024quantitative, breen2024generative}.

Several approaches have been suggested to overcome the challenge set by the unwanted color variation. A simple yet logistically non-trivial solution is to source training data from a large number of different centers, which has shown to improve the generalizability of the AI models \cite{therrien2018role}. However, ensuring  a representative range of variation is a big challenge. Multi-center setups are typically limited in breadth, with source data commonly originating from 2 to 6 sites. Such setups hardly capture the full range of variation across laboratories. Another approach to increase variation in training data is  synthetic data augmentation which is commonly used in machine learning as  a way to mimic the variations by computational means. It helps in reducing generalization errors \cite{tellez2019quantifying}. The augmentation techniques, however, still fall short of simulating real world color and stain variations because of their highly non-linear and dynamic nature and partially owing to  the lack of datasets revealing the full range of variation, hindering the possibility to ground the augmentation in actual observed variation.

An alternative approach to stain augmentation is to reduce the variation in the input data originating from different data domains, such as the staining differences between source data centers. This approach is called stain normalization, where tissue images from one center are normalized with respect to a representative image or set of images from another center. Like data normalization in general, it can help to improve model generalizability, convergence pace, and robustness against overfitting to the extreme ends of the color spectrum. Stain normalization not only helps AI models to perform well with data originating from different sources, but a recent study has shown that the color uniformity of stain normalized tissue images can even help pathologists to speed up diagnoses with increased diagnostic confidence \cite{salvi2023impact}. Stain normalization methods are mainly divided into two categories: traditional methods typically based on mathematical frameworks and AI-based methods \cite{breen2024generative} mostly using generative adversarial network  (GAN) \cite{goodfellow2014generative}.

Traditional stain normalization methods based on mathematical frameworks often employ techniques such as moments (statistical) matching in different color spaces or color deconvolution whereby stains are first decoupled and then normalized separately. Typically, traditional methods rely on a reference (representative) patch to learn the target staining template and the choice of the reference patch is subjective depending on a specialist's assessment. In some cases, a composite patch is generated by combining several patches containing different morphological elements of the tissue. Some of the most commonly used of such approaches are Reinhard \cite{reinhard2001color}, Macenko \cite{macenko2009method}, Vahadane \cite{vahadane2016structure}, and the color histogram matching \cite{gonzalez2008digital}. Prior to the widespread adoption of deep learning, although not commonplace, traditional stain normalization techniques were mostly used as a standardization step in tissue analysis \cite{magee2009colour, niethammer2010appearance, fuchs2011computational}.

AI-based stain normalization predominantly employs GANs \cite{goodfellow2014generative} as an image-to-image translation method to transform source or unnormalized tissue images to match the color of the target or reference tissue image or images. Generative models are typically used to generate synthetic data. A typical GAN consists of a generator model, responsible for data synthesis, and a discriminator model, responsible for quality assurance feedback of the synthesized data. Both components are trained in a game theoretic way until a performance equilibrium is reached. In AI-based stain normalization, both unsupervised and supervised learning approaches are employed. CycleGAN \cite{zhu2017unpaired}, a very commonly used unsupervised image-to-image translation technique that does not rely on aligned image pairs, has been used for stain-to-stain translation \cite{levy2020preliminary, gadermayr2018way} and virtual histopathology staining \cite{koivukoski2023unstained, khan2023effect, salido2023comparison}. Similarly, it has been very effective for stain normalization as well, de Bel et al. used CycleGAN, with a UNet-like \cite{ronneberger2015u} generator architecture, to normalize cross-center whole slide image (WSI) data consisting of Periodic Acid-Schiff (PAS) stained renal tissue. They further tested the effectiveness of normalization on a downstream segmentation task and observed a significant increase in the Dice coefficient of normalized WSIs \cite{de2018stain}. De Bel et al. again used CycleGAN, with enhanced configuration using residual learning, on WSI data sourced from six different centers with a mix of colon and kidney samples stained with H\&E and PAS. They demonstrated that the modified CycleGAN with residual learning outperforms traditional methods and vanilla CycleGAN on the downstream segmentation task keeping the tissue structure intact \cite{de2021residual}. A supervised image-to-image translation method called Pix2pix \cite{isola2017image} has been widely adopted in domains where aligned image data is available. Despite aligned image pairs not being a possibility in the case of stain normalization, for supervised learning methods, typically, the input-output image pair consists of the grayscale version of the stained tissue patch as the input, and the RGB version of the same patch as the output which are inherently aligned. Salehi et al. used Pix2pix with such experimental setup and using quantitative and perceptual evaluation demonstrated the effectiveness of this method over traditional stain normalization approaches \cite{salehi2020pix2pix}.

In recent years, many similar GAN-based methods have been proposed for stain normalization \cite{shaban2019staingan, cho1710neural, bentaieb2017adversarial, cong2021semi, breen2024generative}. While these studies provide valuable information about the strengths of these methods, their scope is limited by the number of centers involved. Most studies sourced tissue image data from two sites with a few exceptions, such as De Bal et al. \cite{de2021residual}, wherein six different sites provided tissue images for the study. In this study, we addressed the issue by analyzing the breadth of staining variation and compared different traditional and GAN-based stain normalization methods using a tissue image dataset with an unprecedented extent of variation, composed of H\&E-stained slides collected from 66 different laboratories across 11 countries. We further provide the dataset as a resource for further stain normalization testing and development.

\section*{Results}
\begin{figure}[htp!]
\centering
\includegraphics[width=\linewidth]{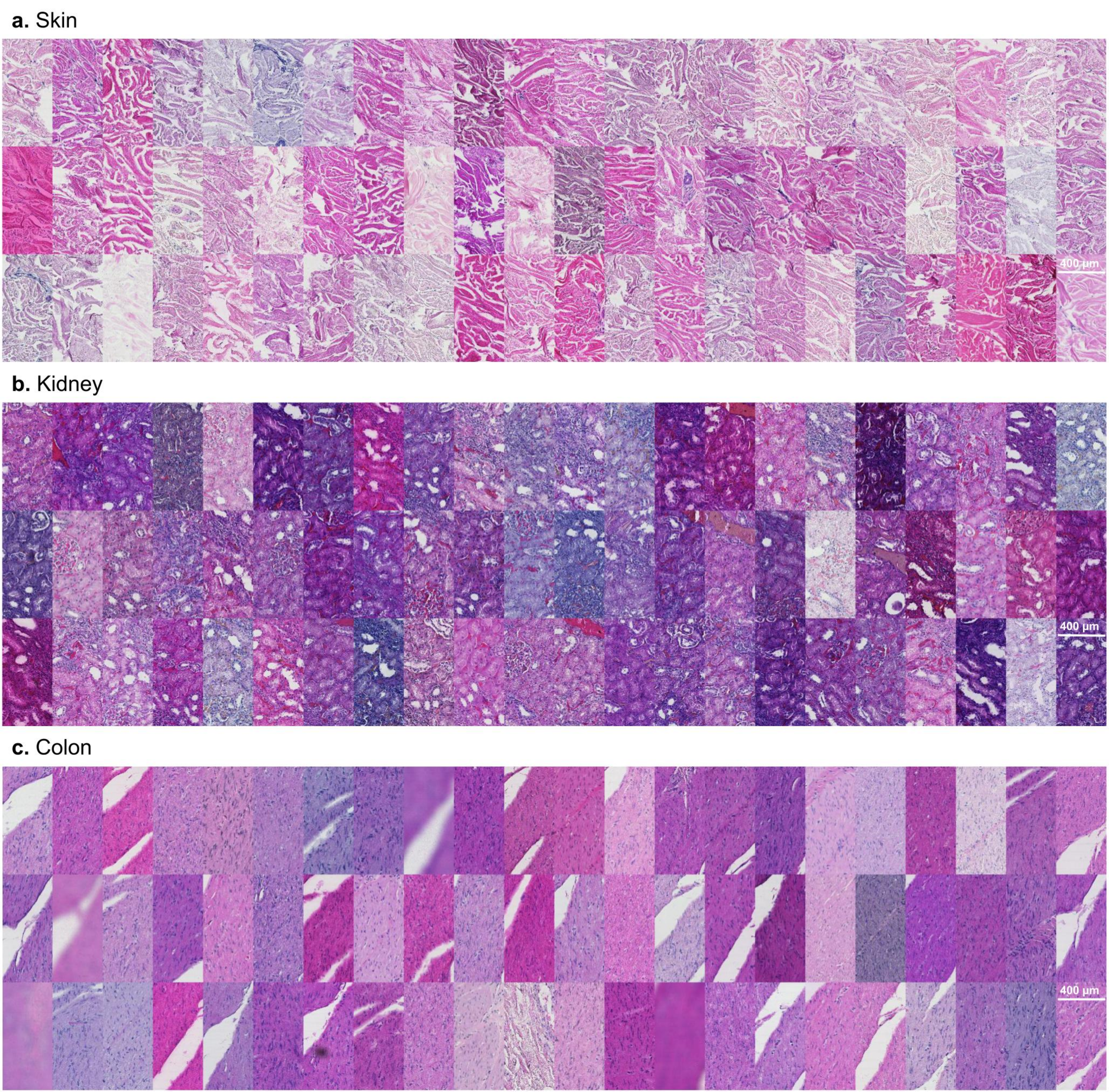}
\caption{A collage of tissue patches representing variance in appearance. \\
\textbf{a.} Image patches extracted from the dermis layer of skin tissue sections. \textbf{b.} Image patches extracted from kidney tissue sections. \textbf{c.} Image patches extracted from the smooth muscle layer of colon tissue sections*.}
\label{fig:variance}
\end{figure}

\begin{figure}[htp!]
\centering
\includegraphics[width=\linewidth]{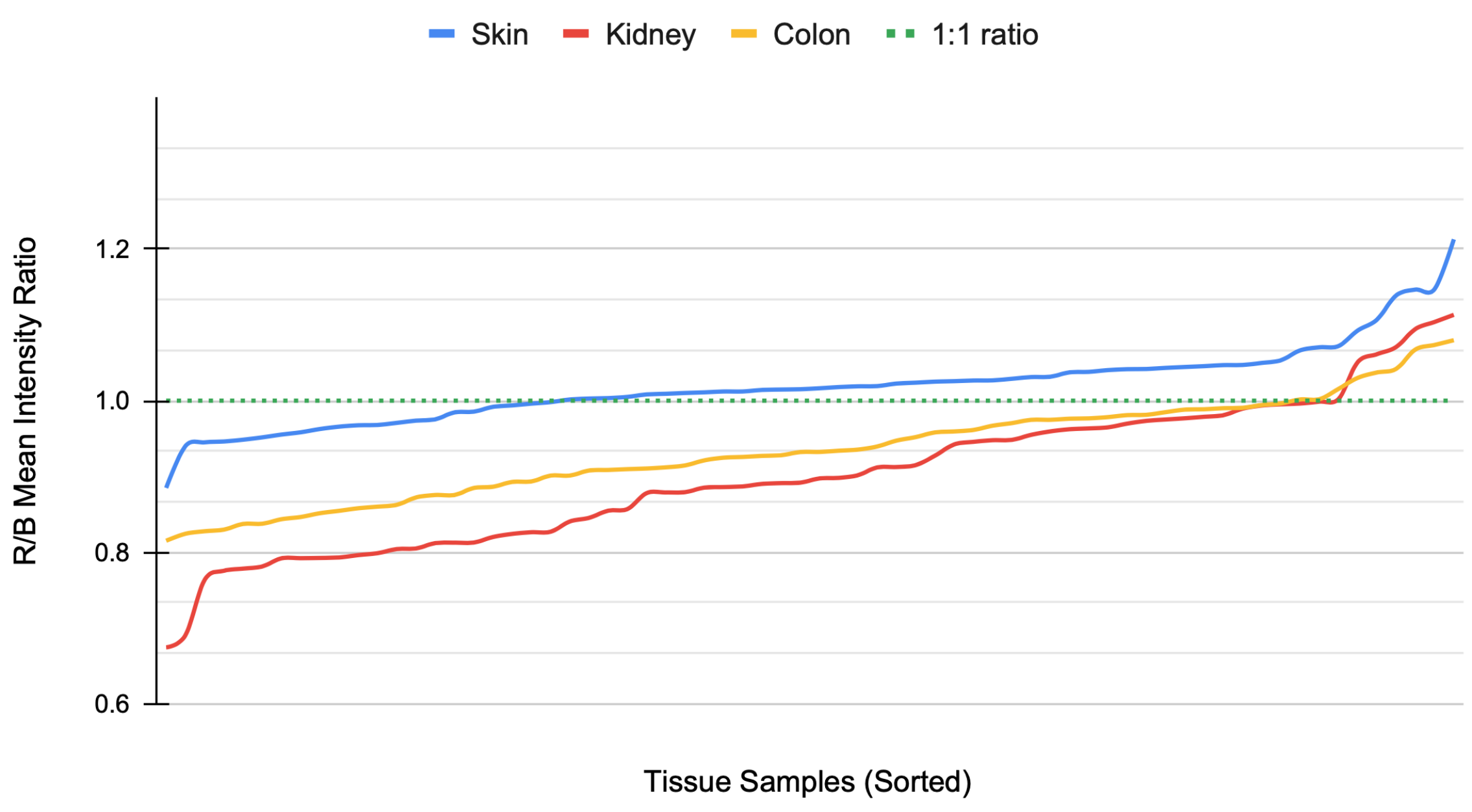}
\caption{Pre-normalization red and blue channel mean intensity plot. \\
The plot shows the spread of color variation based on the ratio of red and blue channel mean intensities of the tissue images. All three tissues, i.e., skin, kidney and colon WSIs deviate drastically from the color-balanced (dotted) line.}
\label{fig:variance-plot}
\end{figure}

The image data used in this study consists of H\&E-stained skin, kidney, and colon tissue sections originating from the same blocks but stained in different laboratories. As expected, the visual appearance of the stained tissue samples varied quite drastically due to color variation originating from different stain compositions and staining protocols in use at different sites (Figure \ref{fig:variance}). To examine the differences in color appearance quantitatively, we computed the mean intensity of red and blue channels and plotted the ratio of all the samples (Figure \ref{fig:variance-plot}). We then performed a comparative analysis of four traditional stain normalization methods, i.e., histogram matching \cite{gonzalez2008digital}, Macenko \cite{macenko2009method}, Reinhard \cite{reinhard2001color}, and Vahadane \cite{vahadane2016structure} and two GAN-based methods CycleGAN \cite{zhu2017unpaired} and Pix2pix \cite{isola2017image} with two variants each. For CycleGAN two different generators were used, one with a UNet-based \cite{ronneberger2015u} generator and the other with a ResNet-based \cite{he2016deep} generator. For Pix2pix, one with UNet-based generator and the other with DenseUNet-based generator which has shown to significantly reduce hallucination artifacts in a relatively similar histological image-to-image translation problem, i.e., virtual H\&E staining of unstained tissue images \cite{khan2023effect}.

\subsection*{Quantitative Evaluation}
We used three different types of quantitative evaluation methods to assess the performance of stain normalization methods. First, to assess the performance on color transfer task, the normalized images were transformed to the $l\alpha\beta$ color space \cite{international2004image} and then the normalized histogram of each channel was compared with the reference image histograms using intersection, Pearson correlation coefficient (PCC), euclidean distance,  and Jensen-Shannon (JS) divergence. Then, using an InceptionV3 model \cite{szegedy2016rethinking}, bottleneck features were extracted for all normalized images and were compared against reference image features using Frachet Inception Distance (FID) \cite{heusel2017gans}, a distance measure that works on high-level abstracted features taking into account both style and structure. Finally, to evaluate the structural integrity of the normalized tissue images with respect to the original ones, we used structural similarity index measure (SSIM) \cite{zhang2011fsim}. 

\subsubsection*{Skin}
\begin{figure}[htp!]
\centering
\includegraphics[scale=0.2]{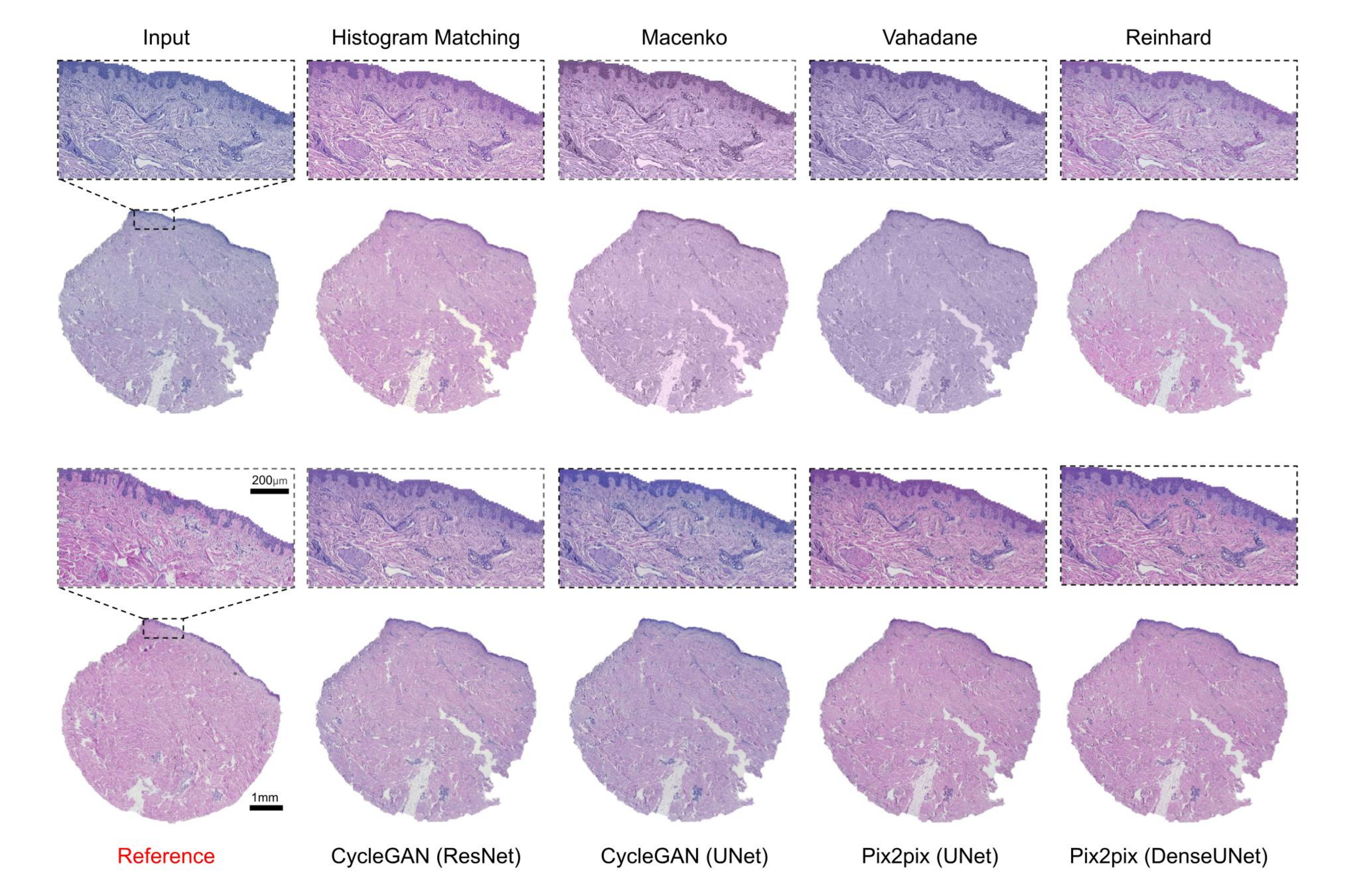}
\caption{Skin tissue stain normalization. \\
Stain-normalized tissue images from eight different methods, i.e., histogram matching, Macenko, Vahadane, Reinhard, CycleGAN (ResNet), CycleGAN (UNet), Pix2pix (UNet), Pixpix (DenseUNet) along with input and reference WSI.}
\label{fig:skin}
\end{figure}

Skin samples were from tissue microarray punch biopsy containing epidermis and dermis (Figure \ref{fig:variance}a). Histogram matching emerged as the most efficient normalization method and achieved mean scores of 0.891, 0.938, 0.279, and 0.119 for intersection, PCC, Euclidean distance and JS divergence, respectively (Table \ref{tab:skin-tab}). As far as high-level feature-based similarity is concerned, histogram matching again, outperformed the rest of the methods with a mean FID score of 61.67 as compared to the original images with a mean FID score of 69.49 (Table \ref{tab:fid-tab}). The structural similarity was best preserved by Vahadane normalization with a mean SSIM score of 0.995, however, it was the worst performing method of all as far as the other metrics are concerned. Methods like Reinhard and histogram matching also had high mean SSIM scores of 0.977 and 0.955, respectively (Table \ref{tab:ssim-tab}). Overall, mean SSIM scores were over 0.92 which shows that tissue structure was mostly preserved well by all the methods (Table \ref{tab:ssim-tab}). Figure \ref{fig:skin} shows an example of stain-normalized skin tissue from each method.

\subsubsection*{Kidney}
\begin{figure}[htp!]
\centering
\includegraphics[scale=0.2]{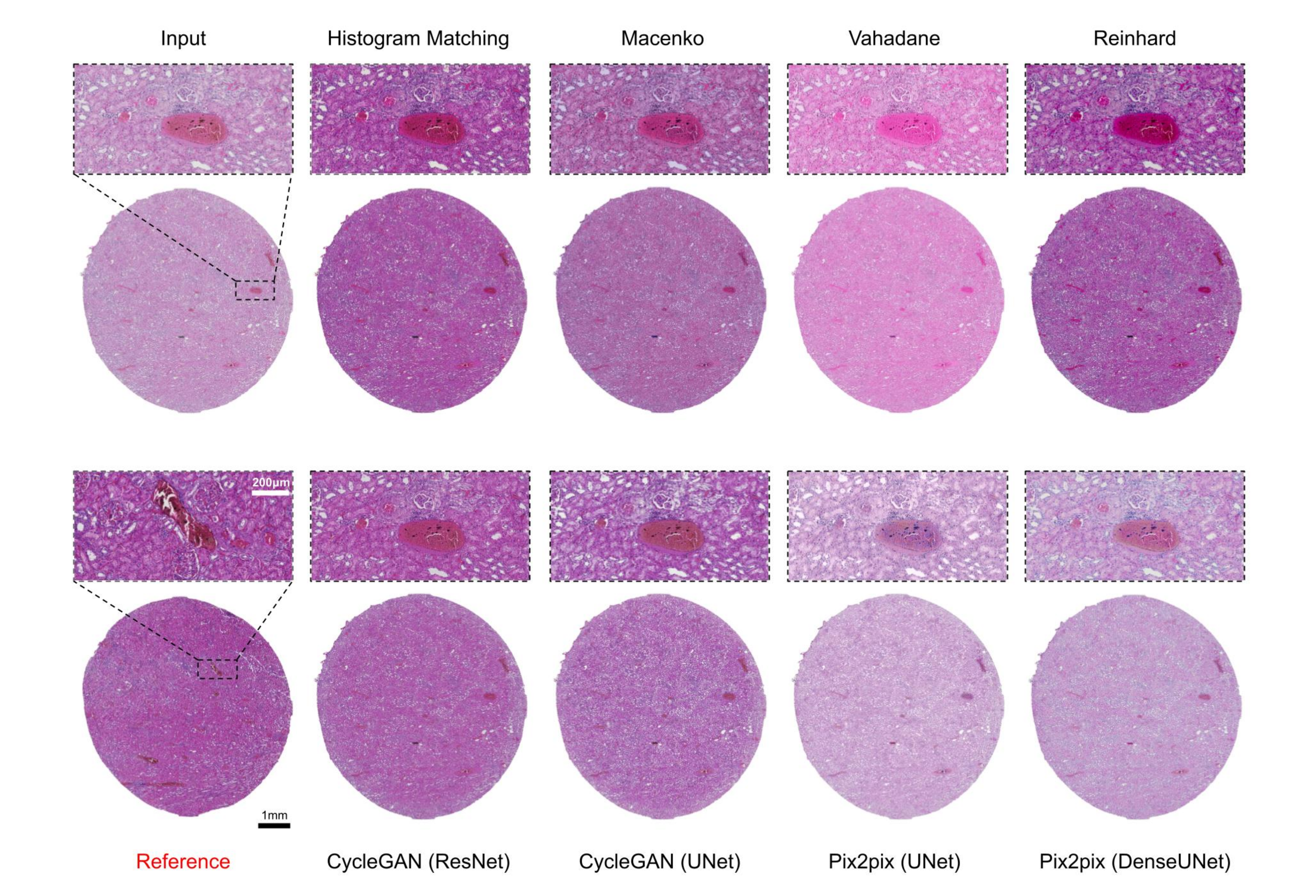}
\caption{Kidney tissue stain normalization. \\
Stain-normalized tissue images from eight different methods, i.e., histogram matching, Macenko, Vahadane, Reinhard, CycleGAN (ResNet), CycleGAN (UNet), Pix2pix (UNet), Pixpix (DenseUNet) along with input and reference WSI. }
\label{fig:kidney}
\end{figure}

Kidney samples were from tissue microarray punch biopsy, representing the cortical layer of the kidney, comprising tubules and renal corpuscles and vasculature (Figure \ref{fig:variance}b). Similar to skin, histogram matching achieved the best mean scores in all the metrics for kidney tissue as well, 0.944, 0.985, 0.144, and 0.101 for intersection, PCC, Euclidean distance and JS divergence, respectively (Table \ref{tab:kidney-tab}). Histogram normalization performed exceptionally well as compared to other methods likely due to the morphological uniformity of kidney tissue mages. In high-level feature comparison, histogram normalization achieved a mean FID score of 55.38 as compared to 69.54 of the original images (Table \ref{tab:ssim-tab}). For SSIM scores we observed a similar trend as skin tissue images, i.e., Vahadane preserving the structural similarity best with a mean SSIM score of 0.967 and all mean SSIM scores being over 0.945 indicating that all the methods performed quite well in preserving the tissue structure (Table \ref{tab:ssim-tab}). Figure \ref{fig:kidney} shows an example of stain-normalized kidney tissue from each method.

\subsubsection*{Colon}
Colon tissue samples comprehensively represented the large intestine, containing mucosa, muscularis mucosae, submucosa and the outer muscle layers, with occasional mucosa-associated lymphoid structures (Figure \ref{fig:variance}c). Intra-tissue morphological heterogeneity peaked in colon tissue as compared to the skin and kidney tissues, however, since the tissue specimens were sectioned from the same tissue block there still existed inter-tissue structural uniformity. We noticed that even for colon tissues images, histogram matching outperformed other methods with a mean scores of 0.906, 0.935, 0.295, and 0.143 for intersection, PCC, Euclidean distance and JS divergence, respectively (Table \ref{tab:colon-tab}). In high-level feature comparison, however, Reinhard and both CycleGAN (Resnet-based and Unet-based) variants performed better than histogram matching with the mean FID scores of 92.37, 96.12, 96.50, and 99.92, respectively (Table \ref{tab:fid-tab}). Similar to skin and kidney, Vahadane performed the best in preserving structural similarity with a mean SSIM score of 0.989 (Table \ref{tab:ssim-tab}). It was closely followed by Reinhard and histogram matching with mean SSIM scores of 0.968 and 0.951 (Table \ref{tab:ssim-tab}). Figure 4 shows an example of stain-normalized colon tissue from each method.

\begin{table}[htp!]
\centering
\begin{tabular}{|l|cc|cc|cc|cc|}
\hline
\multirow{2}{*}{\textbf{Methods}} & \multicolumn{2}{c|}{\textbf{Intersection}} & \multicolumn{2}{c|}{\textbf{PCC}} & \multicolumn{2}{c|}{\textbf{Euclidean Distance}} & \multicolumn{2}{c|}{\textbf{JS Divergence}} \\
\cline{2-9}
 & Mean & Std & Mean & Std & Mean & Std & Mean & Std \\
\hline
CycleGAN (Resnet) & 0.871 & 0.042 & 0.926 & 0.050 & 0.321 & 0.097 & 0.140 & 0.050 \\
\hline
CycleGAN (UNet) & 0.840 & 0.050 & 0.885 & 0.063 & 0.398 & 0.111 & 0.167 & 0.062 \\
\hline
Pix2pix (UNet) & 0.816 & 0.071 & 0.851 & 0.099 & 0.436 & 0.150 & 0.202 & 0.077 \\
\hline
Pix2pix (DenseUNet) & 0.822 & 0.077 & 0.856 & 0.104 & 0.425 & 0.161 & 0.191 & 0.080 \\
\hline
\textbf{Histogram Matching} & \textbf{0.891} & 0.034 & \textbf{0.938} & 0.050 & \textbf{0.279} & 0.101 & \textbf{0.119} & 0.041 \\
\hline
Macenko & 0.822 & 0.055 & 0.748 & 0.095 & 0.612 & 0.119 & 0.240 & 0.058 \\
\hline
Reinhard & 0.829 & 0.062 & 0.819 & 0.102 & 0.499 & 0.147 & 0.207 & 0.079 \\
\hline
Vahadane & 0.706 & 0.056 & 0.638 & 0.092 & 0.732 & 0.102 & 0.345 & 0.054 \\
\hline
Original & 0.779 & 0.088 & 0.803 & 0.119 & 0.503 & 0.170 & 0.220 & 0.081 \\
\hline
\end{tabular}
\caption{\label{tab:skin-tab}Skin tissue samples stain normalization comparative evaluation stats. \\
Table shows mean intersection, Pearson correlation coefficient (PCC), Euclidean distance, Jensen-Shannon (JS) divergence scores of color-normalized WSIs, generated by the eight methods and the original WSIs compared with the reference WSI. Scores were computed on channel histograms in the $l\alpha\beta$ color space. Bold digits represent the best performing method.}
\end{table}

\begin{table}[htp!]
\centering
\begin{tabular}{|l|cc|cc|cc|cc|}
\hline
\multirow{2}{*}{\textbf{Methods}} & \multicolumn{2}{c|}{\textbf{Intersection}} & \multicolumn{2}{c|}{\textbf{PCC}} & \multicolumn{2}{c|}{\textbf{Euclidean Distance}} & \multicolumn{2}{c|}{\textbf{JS Divergence}} \\
\cline{2-9}
 & Mean & Std & Mean & Std & Mean & Std & Mean & Std \\
\hline
CycleGAN (Resnet) & 0.895 & 0.040 & 0.927 & 0.040 & 0.310 & 0.070 & 0.159 & 0.027 \\
\hline
CycleGAN (UNet) & 0.857 & 0.062 & 0.890 & 0.089 & 0.370 & 0.108 & 0.187 & 0.042 \\
\hline
Pix2pix (UNet) & 0.782 & 0.139 & 0.764 & 0.217 & 0.515 & 0.214 & 0.246 & 0.101 \\
\hline
Pix2pix (DenseUNet) & 0.794 & 0.121 & 0.795 & 0.188 & 0.472 & 0.191 & 0.227 & 0.085 \\
\hline
\textbf{Histogram Matching} & \textbf{0.944} & 0.022 & \textbf{0.985} & 0.008 & \textbf{0.144} & 0.037 & \textbf{0.101} & 0.030 \\
\hline
Macenko & 0.784 & 0.065 & 0.823 & 0.093 & 0.476 & 0.120 & 0.242 & 0.045 \\
\hline
Reinhard & 0.854 & 0.072 & 0.885 & 0.075 & 0.389 & 0.106 & 0.195 & 0.053 \\
\hline
Vahadane & 0.676 & 0.078 & 0.578 & 0.135 & 0.770 & 0.140 & 0.378 & 0.056 \\
\hline
Original & 0.721 & 0.115 & 0.722 & 0.189 & 0.549 & 0.200 & 0.251 & 0.093 \\
\hline
\end{tabular}
\caption{\label{tab:kidney-tab}Kidney tissue samples stain normalization comparative evaluation stats. \\
Table shows mean intersection, Pearson correlation coefficient (PCC), Euclidean distance, Jensen-Shannon (JS) divergence scores of color-normalized WSIs, generated by the eight methods and the original WSIs compared with the reference WSI. Scores were computed on channel histograms in the $l\alpha\beta$ color space. Bold digits represent the best performing method.}
\end{table}

\begin{figure}[htp!]
\centering
\includegraphics[scale=0.2]{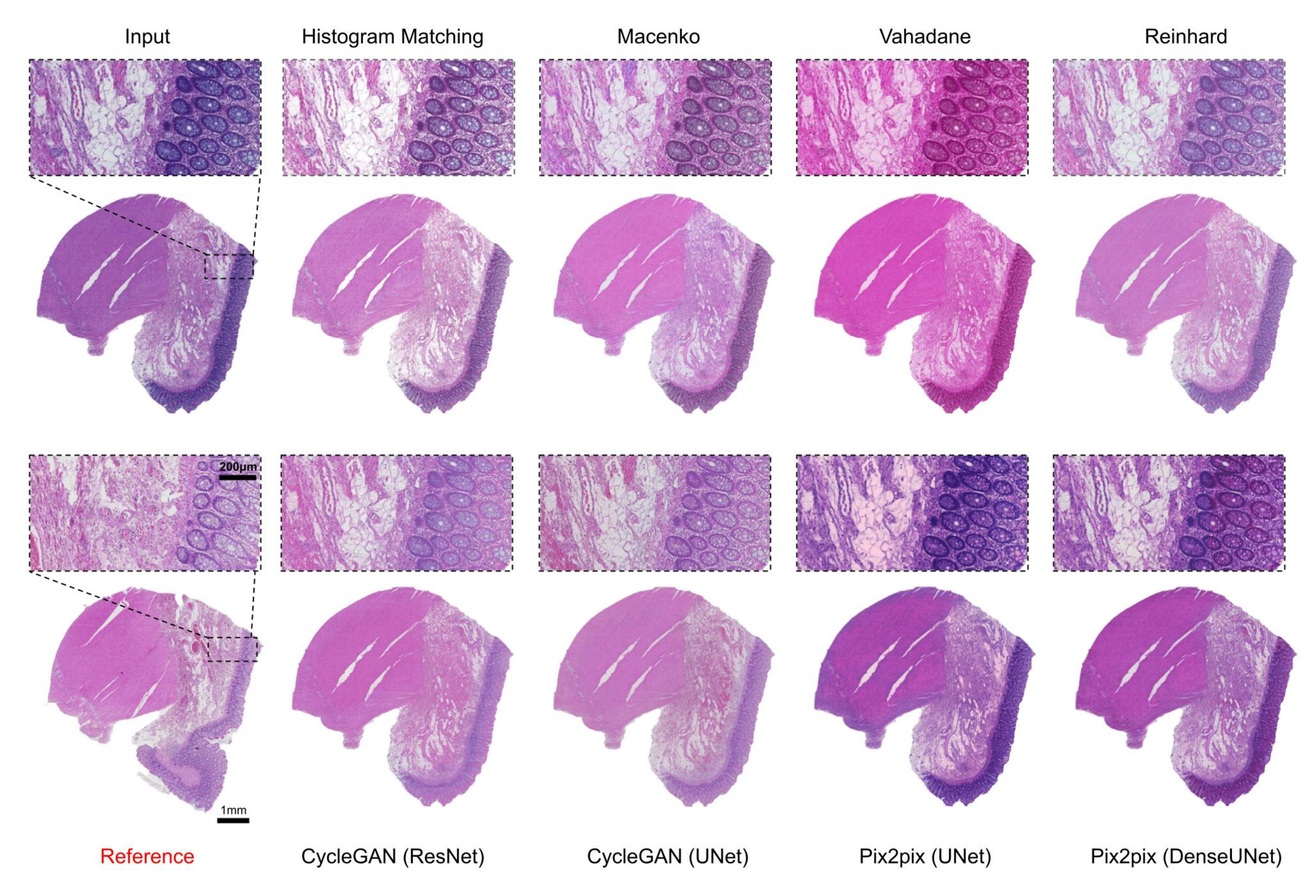}
\caption{Colon tissue stain normalization. \\
Stain-normalized tissue images from eight different methods, i.e., histogram matching, Macenko, Vahadane, Reinhard, CycleGAN (ResNet), CycleGAN (UNet), Pix2pix (UNet), Pixpix (DenseUNet) along with input and reference WSI.}
\label{fig:colon}
\end{figure}

\begin{table}[htp!]
\centering
\begin{tabular}{|l|cc|cc|cc|cc|}
\hline
\multirow{2}{*}{\textbf{Methods}} & \multicolumn{2}{c|}{\textbf{Intersection}} & \multicolumn{2}{c|}{\textbf{PCC}} & \multicolumn{2}{c|}{\textbf{Euclidean Distance}} & \multicolumn{2}{c|}{\textbf{JS Divergence}} \\
\cline{2-9}
 & Mean & Std & Mean & Std & Mean & Std & Mean & Std \\
\hline
CycleGAN (Resnet) & 0.869 & 0.045 & 0.875 & 0.045 & 0.422 & 0.079 & 0.186 & 0.035 \\
\hline
CycleGAN (UNet) & 0.803 & 0.091 & 0.800 & 0.113 & 0.518 & 0.149 & 0.231 & 0.070 \\
\hline
Pix2pix (UNet) & 0.727 & 0.100 & 0.667 & 0.178 & 0.630 & 0.178 & 0.279 & 0.078 \\
\hline
Pix2pix (DenseUNet) & 0.748 & 0.094 & 0.681 & 0.158 & 0.633 & 0.164 & 0.270 & 0.067 \\
\hline
\textbf{Histogram Matching} & \textbf{0.906} & 0.022 & \textbf{0.935} & 0.026 & \textbf{0.295} & 0.058 & \textbf{0.143} & 0.035 \\
\hline
Macenko & 0.817 & 0.049 & 0.810 & 0.072 & 0.522 & 0.094 & 0.246 & 0.042 \\
\hline
Reinhard & 0.839 & 0.061 & 0.839 & 0.074 & 0.477 & 0.108 & 0.223 & 0.063 \\
\hline
Vahadane & 0.726 & 0.066 & 0.662 & 0.120 & 0.695 & 0.127 & 0.336 & 0.055 \\
\hline
Original & 0.739 & 0.082 & 0.705 & 0.131 & 0.595 & 0.153 & 0.260 & 0.069 \\
\hline
\end{tabular}
\caption{\label{tab:colon-tab}Colon tissue samples stain normalization comparative evaluation stats. \\
Table shows mean intersection, Pearson correlation coefficient (PCC), Euclidean distance, Jensen-Shannon (JS) divergence scores of color-normalized WSIs, generated by the eight methods and the original WSIs compared with the reference WSI. Scores were computed on channel histograms in the $l\alpha\beta$ color space. Bold digits represent the best performing method.}
\end{table}

\begin{table}[htp!]
\centering
\begin{tabular}{|l|cc|cc|cc|}
\hline
\multirow{2}{*}{\textbf{Methods}} & \multicolumn{2}{c|}{\textbf{Skin}} & \multicolumn{2}{c|}{\textbf{Kidney}} & \multicolumn{2}{c|}{\textbf{Colon}} \\
\cline{2-7}
& Mean & Std & Mean & Std & Mean & Std \\
\hline
CycleGAN (Resnet) & 92.54 & 20.57 & 96.27 & 21.38 & 96.12 & 40.40 \\
\hline
CycleGAN (UNet) & 101.63 & 21.57 & 91.68 & 24.22 & 96.50 & 18.54 \\
\hline
Pix2pix (UNet) & 109.00 & 20.49 & 118.17 & 35.51 & 123.10 & 54.75 \\
\hline
Pix2pix (DenseUNet) & 109.06 & 22.67 & 117.72 & 36.21 & 119.05 & 55.92 \\
\hline
Histogram Matching & \textbf{61.67} & 24.29 & \textbf{55.38} & 38.62 & 99.92 & 61.76 \\
\hline
Macenko & 89.65 & 40.54 & 106.51 & 52.78 & 134.63 & 52.69 \\
\hline
Reinhard & 62.08 & 31.30 & 64.43 & 39.68 & \textbf{92.37} & 65.58 \\
\hline
Vahadane & 134.57 & 17.99 & 153.15 & 49.35 & 150.15 & 55.01 \\
\hline
Original & 69.49 & 37.35 & 69.54 & 42.26 & 96.88 & 64.28 \\
\hline
\end{tabular}
\caption{\label{tab:fid-tab}Frachet inception distance (FID) for skin, kidney and colon tissue samples. \\
Color-normalized WSIs, generated by the eight methods and the original WSIs were compared using FID, a metric that uses high-level features extracted through the bottleneck layer of models like InceptionV3. Bold digits represent the best performing method.}
\end{table}

\begin{table}[htp!]
\centering
\begin{tabular}{|l|cc|cc|cc|}
\hline
\multirow{2}{*}{\textbf{Methods}} & \multicolumn{2}{c|}{\textbf{Skin}} & \multicolumn{2}{c|}{\textbf{Kidney}} & \multicolumn{2}{c|}{\textbf{Colon}} \\
\cline{2-7}
& Mean & Std & Mean & Std & Mean & Std \\
\hline
CycleGAN (Resnet) & 0.940 & 0.019 & 0.946 & 0.023 & 0.860 & 0.117 \\
\hline
CycleGAN (UNet) & 0.920 & 0.024 & 0.949 & 0.015 & 0.822 & 0.118 \\
\hline
Pix2pix (UNet) & 0.955 & 0.010 & 0.957 & 0.019 & 0.933 & 0.015 \\
\hline
Pix2pix (DenseUNet) & 0.958 & 0.010 & 0.963 & 0.008 & 0.945 & 0.011 \\
\hline
Histogram Matching & 0.955 & 0.048 & 0.949 & 0.050 & 0.951 & 0.040 \\
\hline
Macenko & 0.937 & 0.067 & 0.945 & 0.034 & 0.891 & 0.039 \\
\hline
Reinhard & 0.977 & 0.027 & 0.956 & 0.045 & 0.968 & 0.033 \\
\hline
Vahadane & \textbf{0.995} & 0.005 & \textbf{0.967} & 0.027 & \textbf{0.989} & 0.011 \\
\hline
\end{tabular}
\caption{\label{tab:ssim-tab}Structural similarity index measure (SSIM) for skin, kidney and colon tissue samples. \\
Color-normalized WSIs, generated by the eight methods, were compared against the original WSIs to test the (tissue) structural preservation of the methods, using structural similarity index measure (SSIM). Bold digits represent the best performing method.}
\end{table}

\subsection*{Qualitative Evaluation}
We examined the quality of the normalized staining by visually comparing a subset of original and stain-normalized colon specimens, with respect to distinguishability of tissue structures. We evaluated the visual quality of hematoxylin-eosin hue and contrast, as well as structural or other digital artifacts, as these factors contribute to the interpretability of histopathological H\&E images. Since the colon contains several different tissue structures and types, it serves as a representative example that we especially focus here on.  The colon consists of four distinct layers: a) the mucosa, comprising glandular epithelia with interconnected connective tissue (lamina propria) and underlying smooth muscle (muscularis mucosae); b) the submucosa, comprising dense connective tissue with blood and lymphatic vessels; c) muscularis externa composed of smooth muscle layers; as well as d) serosa or adventitia comprising thin connective tissue and an outer layer of squamous epithelia or mesothelium. As the inner two layers include prominent variation in structure, harboring epithelial cells, fibroblasts, vessels, erythrocytes, and smooth muscle, we primarily focused on these layers, while asking how well the normalization preserves fine detail. Because over-staining, under-staining as well as uneven staining were the most notable features of the unprocessed tissue  images, we focused on the mitigation of these issues when evaluating the effectiveness of normalization methods. In addition, we compared the methods to a reference section, to evaluate how well the methods achieved the sought-out result.

\subsubsection*{Mitigation of staining quality issues}
\begin{figure}[htp!]
\centering
\includegraphics[scale=0.23]{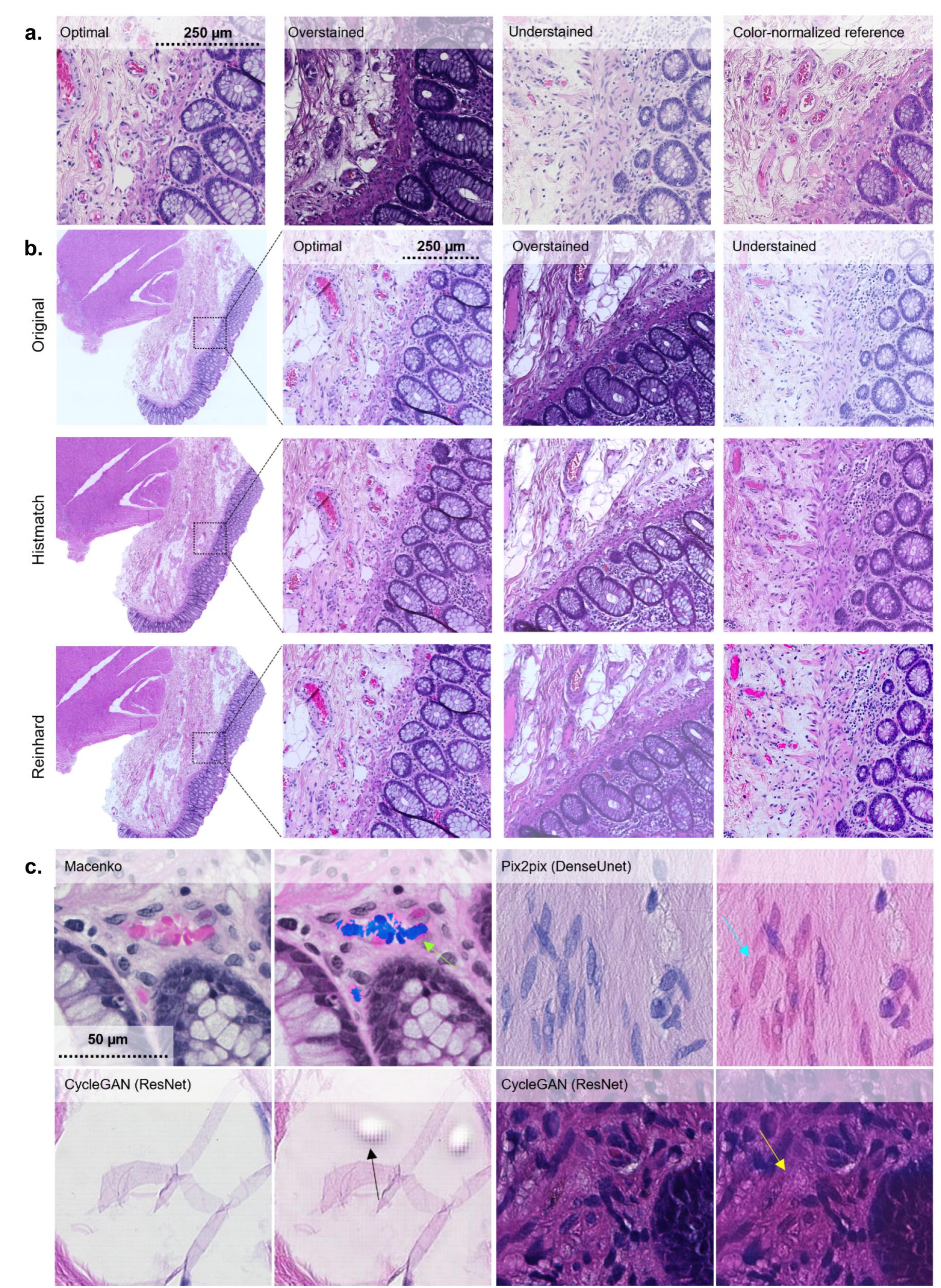}
\caption{Colon tissue qualitative analysis. \\
\textbf{a.} Representative images of optimal as well as over- and understained H\&E-slides. \textbf{b.} Comparisons between original, Histmatch and Reinhard normalized images from colonic mucosa-submucosa border in optimally and suboptimally stained slides. \textbf{c.} Artifacts emerging from image processing.}
\label{fig:qualitative-analysis}
\end{figure}

Visually separable contrast between blue/purple (hematoxylin) and pink/red (eosin), as well as the dynamic range of eosin staining, are central for high-quality H\&E staining suitable for histological interpretation. Thus, the hue of the H\&E-image is an important determinant of quality. A reference image of a high-quality H\&E is shown in Figure \ref{fig:qualitative-analysis}a. With the applied methodology, in general, if normalization resulted in a global shift in hue, it was towards pink/red. A hue-shift of varying strength was present in all methods (Figure \ref{fig:colon}). Most notably, the Vahadane method transformed the white background to pink in most instances, and completely disposed of hematoxylin's blue/purple. The shift towards pink/red may compromise the separability of veins, fibrous components, and actin bundles. With histogram matching, in some instances, the hue was improved with heavily over-stained images, resulting in a more conventional coloration and pronounced contrast, albeit under-stained images tended to shift heavily towards pink/red (Figure \ref{fig:qualitative-analysis}b).  With under-stained images, Macenko and Reinhard performed best in correcting substandard contrast, while not performing well with overstaining (Figure \ref{fig:qualitative-analysis}b). 

Some of the utilized normalization methods generated digital artifacts. From the conventional methods, Macenko transformed the color of erythrocytes and parts of the lamina propria to blue. With deep-learning models, generative artifacts were observed with CycleGAN (ResNet) in empty (fat containing) regions of adipose tissue, as well as, with Pix2pix (DenseUNet), some spindly nuclei from smooth-muscle were transformed red. In addition, CycleGAN (ResNet) incorporated some context-dependent tile artifacts at luminal epithelium – smooth muscle borders (Figure \ref{fig:qualitative-analysis}c). We did not observe artifacts with histogram matching and Reinhard. 

All in all, no method improved visual quality in all instances of suboptimal staining. Histogram matching resulted in improved quality in heavily over-stained images, albeit not performing equally well with under-stained images. Reinhard improved under-stained images, albeit producing below par results in over-stained images. 

\subsubsection*{Performance comparison to reference}
\begin{figure}[htp!]
\centering
\includegraphics[width=\linewidth]{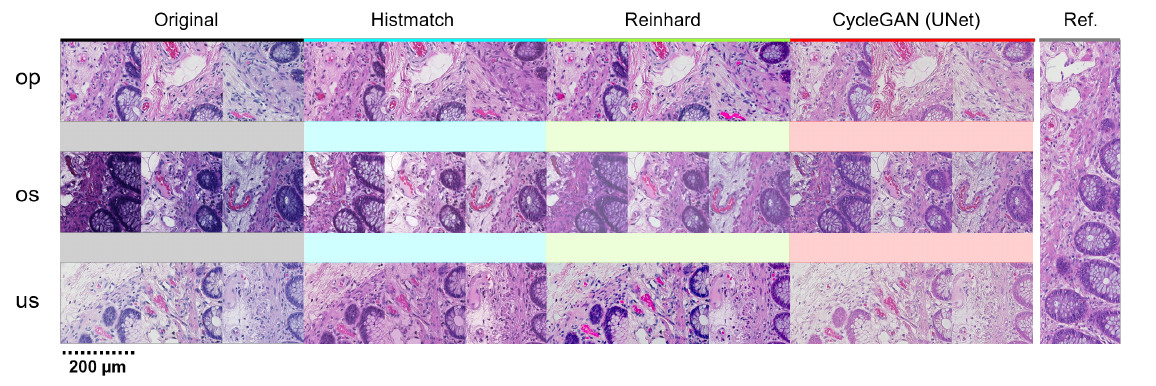}
\caption{Overview of the best performing methods (columns) with varying input staining quality (rows).}
\label{fig:bp-method}
\end{figure}

We next visually evaluated if the methods match the utilized standard, i.e., the color-normalized reference image. The worst performance was observed in Vahadane, which completely discarded hematoxylin. Reinhard increased overall color intensities, resulting in more perceived contrast, thus deviating from the reference. In addition, the hue of hematoxylin varied, and in overstained specimens, the normalized image appeared cloudy. Macenko varied, depending on the original image, most notably in hematoxylin and background hue, as well as blue artifacts in eosin, that deviated from the reference. Macenko also had significant variance and deviation from the reference in both contrast and hue. Reinhard harbored variance in contrast and in the color of the epithelial lining of the lumen. Pix2pix (DenseUNet) was close to reference in most instances, albeit in uneven staining the normalization did deviate. In addition, the other generative AI-models, CycleGAN (UNet) and CycleGan (ResNet) achieved reasonable similarity with the reference. Histogram matching was the best performing method in terms of evenness and comparability to the reference. In histogram matching, the most notable deviation from the reference was the background, which in some instances had a pink hue. However, when considering only the ROI, histogram matching harbored the most even performance compared to a reference image, both with smaller magnification and when evaluating the finer details in different tissue compartments. An overall comparison of the best performing methods is shown in Figure \ref{fig:bp-method}.

\section*{Discussion}
In this work, we present a tissue image dataset consisting of H\&E-stained skin, kidney and colon tissues by staining sections of the same samples in 66 different laboratories to study the variation in the appearance of tissue images and assess the performance of different stain normalization techniques. The tissue sections were cut from the same block for each tissue type, representing a setup where a relatively high level of content similarity can be expected across the 66 sections. Tissue staining is one of the most critical steps in sample preparation and subtle variations in implementation of the staining protocol could contribute to the fluctuation in the appearance of the dyed tissue, commonly observed as variations in the intensity and hue of the staining. Previous studies have mostly employed tissue images sourced from a handful of sites \cite{de2018stain, de2021residual}. This study, however, presents a dataset from numerous laboratories that captures a wide range of variation in H\&E staining (Figure \ref{fig:variance}), while excluding other sources of technical and biological variation. The data is released as an open access resource for the community to enable benchmarking of normalization methods as well as other further investigation of staining variation in histopathology.

The substantial variation in staining offers an opportunity to assess the effectiveness of different strategies to normalize the color heterogeneity across the tissue samples and also evaluate their ability to maintain morphological interpretability of the tissue content. To that end, we employed four traditional and four GAN-based methods to normalize the staining in tissue images. The traditional methods included histogram matching, Macenko, Reinhard, and Vahadane stain normalization. Recently, methods based on generative AI have gained popularity for out-performing traditional methods \cite{de2018stain, de2021residual}. Here, we included GAN-based methods in addition to the traditional normalization methods. The GAN-based methods included two variants of Pix2pix, one with UNet-based generator, the other with DenseUNet-inspired generator, and two variants of CycleGAN, one using UNet-based generator and the other using ResNet-based generator. 

In this study, both quantitative and qualitative evaluation revealed that WSIs normalized by variants of CycleGAN and Pix2pix bore reasonable resemblance with the reference sample and largely maintained structural integrity, however, the methods could not conclusively outperform traditional methods across all tissue types and also produced unwanted hallucination artifacts in some instances. The reason for such performance is potentially rooted in the fact that one WSI per laboratory is insufficient for yielding optimal results from the data-hungry deep learning models. For instance, only the reference WSI and its grayscale version, as a single source-target pair, was used for training both Pix2pix and its DenseUNet variant. Although more than one WSI were used in CycleGAN (both ResNet and UNet variants) trainings, only one representative WSI per cluster (representing smaller subsets of the tissue image dataset) was chosen for the training set for each tissue type. 

\begin{figure}[htp!]
\centering
\includegraphics[width=\linewidth]{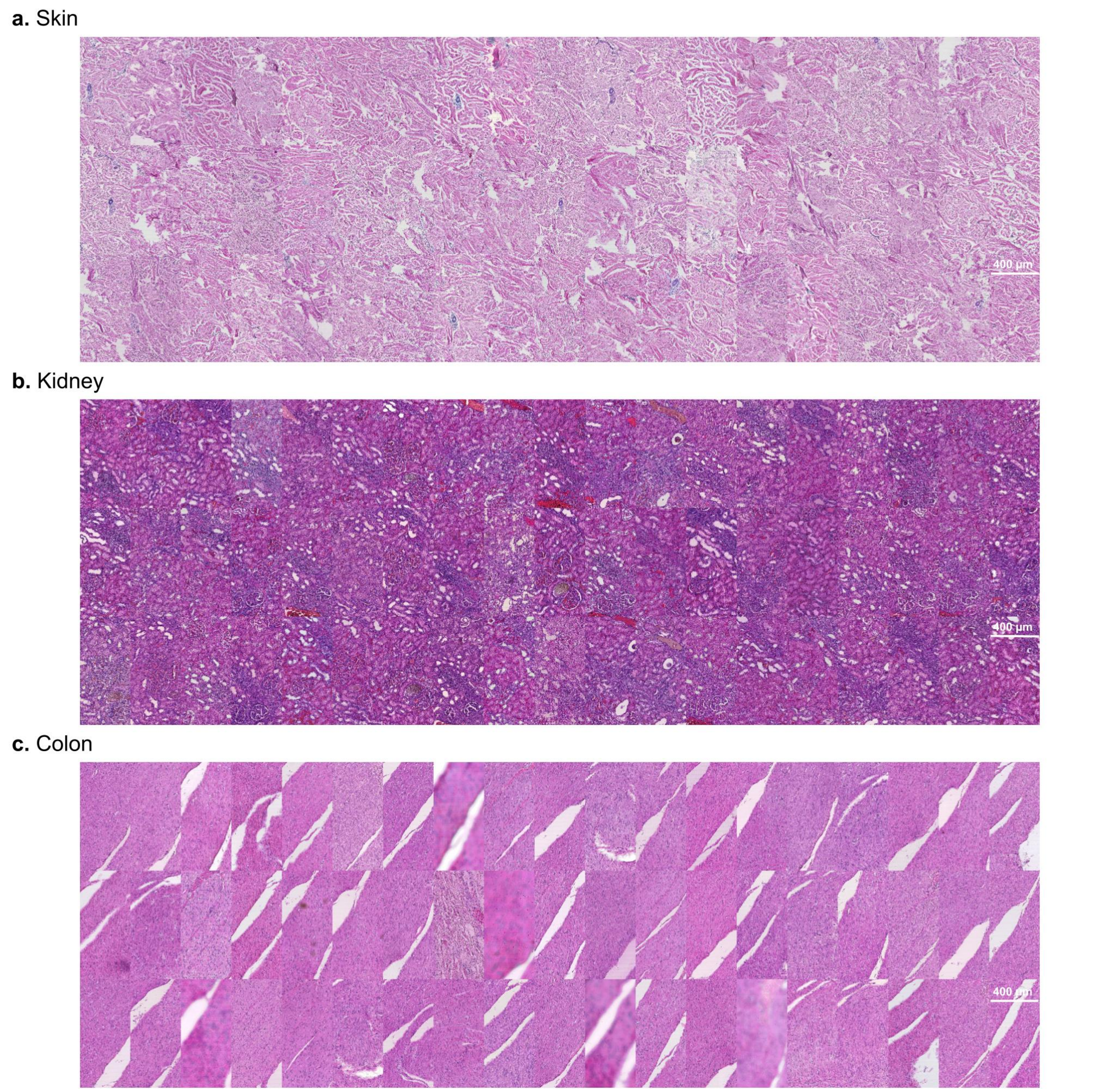}
\caption{Post-normalization tissue patch collage showing the color uniformity. \\
Tissue images stain normalized using histogram normalization. \textbf{a.} Image patches extracted from skin tissue sections*. \textbf{b.} Image patches extracted from kidney tissue sections. \textbf{c.} Image patches extracted from the smooth muscle layer of colon tissue sections.}
\label{fig:normalized}
\end{figure}

\begin{figure}[htp!]
\centering
\includegraphics[width=\linewidth]{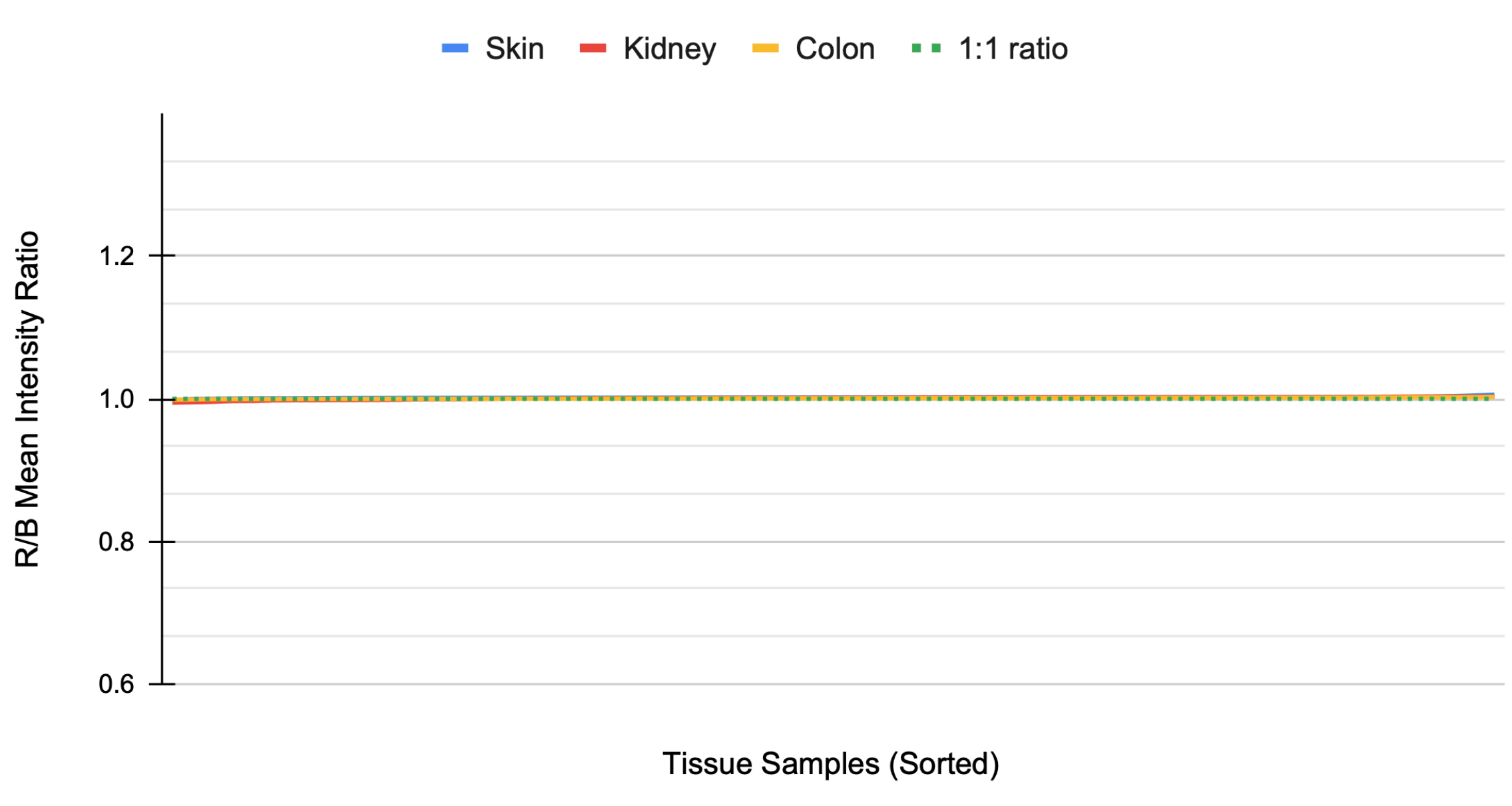}
\caption{Post-normalization red and blue channel mean intensity plot. \\
The plot shows the convergence of red and blue channel mean intensity ratio of the tissue images after histogram matching stain normalization. All three tissue types, i.e., skin, kidney and colon WSIs are virtually on the color-balanced (dotted) line.}
\label{fig:normalized-plot}
\end{figure}

On the other hand, histogram matching as a traditional stain normalization method performed relatively well across all three tissue types compared to all the other methods (Figure \ref{fig:normalized} and \ref{fig:normalized-plot}). This performance can be attributed to at least two different factors. 1) For traditional methods, typically a small representative patch is selected as reference. We, instead, used the whole tissue image as the reference sample, and as opposed to applying normalization patch by patch which typically leads to tiling artifacts, we applied normalization to the WSIs at once. Reinhard and Macenko normalized WSIs also seem to benefit from this strategy. 2) Although the dataset captures a broad spectrum of staining variation, the tissue content across the sections does not change much by virtue of originating from the same tissue block. Further explanations for the better performance of histogram matching and Reinhard normalization could be the fact that both methods adjust the statistical attributes of the image (histogram, mean, and standard deviation) and the adjustments are applied globally, this enables the methods to leverage the morphological similarity factor. However, it is important to note that not all traditional methods yielded satisfactory results. For instance, Macenko sporadically produced blue artifacts in eosin regions, whereas Vahadane had the worst performance as it appeared to infuse a pink hue in the whole tissue image, overriding the hematoxylin component completely. This was clearly reflected in the quantitative evaluation as well.

This study highlights a broad spectrum of variation observed in the inter-laboratory staining manifestation, which also indicates that stain normalization is an indispensable step to homogenize the appearance of the tissue images whenever inter-laboratory comparisons or general tools are used (Figure 6 and 7). Furthermore, it shows that while deep learning-based methods are effective for stain normalization, they have some shortcomings as well, and traditional methods may remain valuable in specific scenarios, such as the present case, where tissue sample morphology exhibits a high degree of uniformity. The dataset is shared as a freely available public resource. In addition to serving as a benchmark for staining normalization methods, it can also benefit the community in determining the extent of staining variation \cite{prezja2022h} which could be leveraged to enhance data augmentation techniques, thereby contributing to the development of more robust AI systems. While the dataset presented here is unique in terms of the variety it showcases, only a single section per laboratory does not facilitate evaluating intra-laboratory variation or quantifying the performance of, e.g., typical deep learning-based stain normalization methods wherein several samples per laboratory are required to effectively train a method and validate its results. A similar dataset containing more samples per laboratory and featuring more morphological variation across tissue samples could provide a stronger foundation for a more comprehensive study in the future.

\section*{Material \& Methods}
\subsection*{Tissue Material}
The H\&E-stained tissue image data was obtained as part of an external quality assessment (EQA) initiative coordinated by Labquality, a company  specializing in external quality assessment programs for clinical laboratories located in Helsinki, Finland. These slides featured a tissue microarray section comprising three 6 mm punch biopsies, each extracted from normal human skin, kidney, and colon. These tissue samples were obtained from anonymized, formalin-fixed, and paraffin-embedded histological specimens from a reference pathology laboratory. During the EQA round, slides with 3-micron thick unstained tissue sections were dispatched to EQA participant laboratories. These laboratories were instructed to apply their routine H\&E staining methodology, typically used in their daily diagnostic practices. In total, 66 laboratories from 11 different countries participated in the evaluation process. Subsequently, these slides were digitized using a Hamamatsu Photonics NanoZoomer-XR slide scanner, employing a 20× objective lens, resulting in a scanning resolution of 0.46 {\textmu}m per pixel.

The tissue images were resampled to 10× to expedite the normalization process, since 10× is sufficient for clinical diagnosis. Another reason for resampling to 10× was to aid traditional normalization method by enabling the use of whole tissue sample as the reference and also apply color normalization to the source tissue at once instead of using just a small tissue patch from the reference tissue sample and applying color normalization patch-wise to the source tissue sample which is the norm with traditional stain normalization methods. The rationale behind this approach was to achieve more uniform results and avoid tiling effects.

\subsection*{Reference Sample Selection}
The reference sample was selected computationally using red-blue channel mean intensity ratio as the criterion. For each tissue type, the sample that had the red-to-blue ratio closest to one was selected as the reference sample (Figure \ref{fig:variance-plot}). The rationale behind selecting the blue and red channels stems from blue and red color expression being characteristic of Hematoxylin and Eosin stained tissue samples.

\subsection*{Traditional Stain Normalization Methods}
\subsubsection*{Histogram Matching}
It is an image processing technique that modifies pixel values in an image to transform its color distributions represented by its histogram so that it matches the color distribution of a reference image \cite{gonzalez2008digital}. By normalizing the color distribution of the source image with respect to the reference image, the process aims to achieve color consistency between the two images, enabling effective image analysis and visualization across varying  imaging conditions. While histogram matching is a general-purpose color normalization technique that is not specifically designed for histopathology images, it has been effectively used in the domain \cite{zewdie2021classification, habtemariam2022cervix}.

\subsubsection*{Macenko}
In Macenko stain normalization, separate H\&E vectors are computed using singular value decomposition (SVD) in the optical density space which is a logarithmic version of the RGB color space. Optical density space makes stain manipulation easier by enabling linear coupling and decoupling of stains. By scaling the stain vectors of the source image to match those of the reference image, Macenko normalizes staining variations, enabling consistent color representation across images. \cite{macenko2009method}

\subsubsection*{Reinhard}
Like histogram normalization, Reinhard normalization \cite{reinhard2001color} is also a generic image normalization technique that was not specifically designed for histopathology images. In this method, the images are first converted to the \textit{l}$\alpha\beta$ color space \cite{international2004image}. Then statistical standardization (subtracting a mean value and scaling by standard deviation) is applied to source images and then the images are linearly scaled using mean and standard deviation of the reference image, this is done separately for each channel. This normalization is effective because unlike RGB, \textit{l}$\alpha\beta$ color space has low correlation between color channels. 

\subsubsection*{Vahadane}
It is another stain normalization approach that employs color deconvolution in optical density space, but unlike Macenko it uses sparse non-negative matrix factorization which is more effective than SVD. The method works by optimizing the deconvolution approach over successive iterations to generate two matrices, one for color appearance and the other for stain density. The source stain density is scaled, a step similar to Macenko’s scaling step, and then by combining with the color appearance of the reference sample a normalized image is reconstructed. \cite{vahadane2016structure}

\noindent We used a publicly available implementation of Macenko, Reinhard, and Vahadane stain normalization methods: \\ \href{https://github.com/wanghao14/Stain_Normalization}{https://github.com/wanghao14/Stain\_Normalization}, whereas histogram matching was used from the Python package skimage.

\subsection*{GAN-based  Stain Normalization Methods}

\subsubsection*{CycleGAN}
One of the most commonly used unsupervised image-to-image translation methods is called CycleGAN. The biggest advantage of using a method like CycleGAN is that it doesn’t necessitate the source and target images to be aligned or have pixel-to-pixel correspondence. The method learns from a mechanism called distribution matching loss, which, in practice, is achieved by using the cycle-consistency loss function in case of CycleGAN. A typical CycleGAN model consists of two generator and two discriminator models. It is capable of bi-directional translation from source to target and vice versa. We used two different variants of CycleGAN generators, one based on ResNet \cite{he2016deep} and the other based on UNet \cite{ronneberger2015u} with skip connections. 

\subsubsection*{Pix2pix}
Domains wherein it is possible to generate paired image data, Pix2pix and its variants are the most widely used GAN-based supervised image-to-image translation methods. It requires the training (image) data to be well registered with pixel-level correspondence because it relies on direct pixel loss functions for its learning. It is based on conditional GAN \cite{isola2017image}. A typical Pix2pix network consists of a pair of generator and discriminator models that are trained in a game-theoretic way until a performance equilibrium is achieved. We used two different variants of Pix2pix generator, one based on UNet \cite{ronneberger2015u} and the other based on DenseUNet \cite{khan2023effect}.

\subsection*{GAN-based Methods Experiment Setup}
While we made it possible to use the whole tissue sample as the so called reference patch for the traditional method, a WSI, as is, can not be fed directly to the GPU for GAN-based model training, therefore, the tissue images were first split into 512px x 512px (433.36{\textmu}m x 433.36{\textmu}m) tiles before the training. 

 CycleGAN training required a representative dataset, since each laboratory provided one tissue per organ, the idea was to treat the whole dataset as one and choose a small representative dataset. We chose a semi-heuristic approach of clustering the tissue images. First, the tissue images were transformed to the \textit{l}$\alpha\beta$ color space \cite{international2004image} for a more independent and distinguishable representation in each channel. Normalized histograms of each channel were used as features. Then dimensionality reduction was applied using principal component analysis, the first two components were used to plot the samples in two dimensions. The components were then used to cluster the samples using k-means clustering \cite{macqueen1967some}. We used the Within-Cluster Sum of Square approach to find the optimal number of clusters. In order to keep the experiments uniform across different organ tissues, we chose 8 clusters and samples closest to the cluster mean were chosen as the representative samples of that cluster. Altogether, 8 samples were chosen to represent the dataset that were used in both CycleGAN trainings. 

 For Pix2pix trainings, each image tile in the reference WSI was first converted to grayscale and then the model was trained on the grayscale and colored tile pairs of the same tissue patch. The objective was to thoroughly train the model to precisely map grayscale tiles to their corresponding color representations. Once the model was trained, the rest of the WSI tiles were fed to the model as grayscale input, to be converted to a similar color as the reference WSI.

\section*{Data and Code availability}
\begin{itemize}
\item The tissue WSI dataset used in this study is freely available on a FAIR-compliant server under \\\href{https://doi.org/10.23729/4cb0b2ed-b074-4e5e-8cd6-ff5242c77fd0}{https://doi.org/10.23729/4cb0b2ed-b074-4e5e-8cd6-ff5242c77fd0}.
\item The implementation of all the stain normalization methods, used in this study, has been deposited at Zenodo under \\ \href{https://doi.org/10.5281/zenodo.12344369}{https://doi.org/10.5281/zenodo.12344369} and are publicly available as of the date of publication. Refer to readme.md for detailed instructions on how to use the code and requirements.txt to install the dependencies.
\end{itemize}

\section*{Acknowledgements}
The authors would like to thank Zuhair Iftikhar and Robin Ekman for their skillful technical assistance in this study. Financial support from Research Council of Finland (PR, LL), Cancer Foundation Finland (PR, LL), Sigrid Juselius Foundation (PR), University of Turku Graduate School (UK) is gratefully acknowledged. Computational experiments were conducted on high-performance computing environment provided by CSC - IT Centre for Science (Finland), the authors are grateful for the resources and support by CSC.

\section*{Author contributions statement}
Conceptualization, P.R., L.L., T.K.; Methodology, U.K., J.H., M.F, L.L., T.K., and P.R.; Software, U.K.; Validation, U.K., J.H., and M.F.; Formal analysis, U.K.; Investigation, U.K., J.H., and M.F.; Resources, T.K. and P.R.; Data Curation, U.K., J.H., and P.R.; Writing – Original Draft, U.K. and J.H.; Writing – Review and Editing, U.K., J.H., L.L., and P.R.; Visualization, U.K., and J.H.; Supervision, T.K. and P.R.; Project Administration, T.K. and P.R.; Funding Acquisition, T.K. and P.R.

\section*{Competing interests}
The authors declare no competing interests.

\newpage
\bibliography{references}

\begin{thebibliography}{10}
\urlstyle{rm}
\expandafter\ifx\csname url\endcsname\relax
  \def\url#1{\texttt{#1}}\fi
\expandafter\ifx\csname urlprefix\endcsname\relax\def\urlprefix{URL }\fi
\expandafter\ifx\csname doiprefix\endcsname\relax\def\doiprefix{DOI: }\fi
\providecommand{\bibinfo}[2]{#2}
\providecommand{\eprint}[2][]{\url{#2}}

\bibitem{troiano2009effects}
\bibinfo{author}{Troiano, N.~W.}, \bibinfo{author}{Ciovacco, W.~A.} \& \bibinfo{author}{Kacena, M.~A.}
\newblock \bibinfo{journal}{\bibinfo{title}{The effects of fixation and dehydration on the histological quality of undecalcified murine bone specimens embedded in methylmethacrylate}}.
\newblock {\emph{\JournalTitle{Journal of histotechnology}}} \textbf{\bibinfo{volume}{32}}, \bibinfo{pages}{27--31} (\bibinfo{year}{2009}).

\bibitem{chlipala2021impact}
\bibinfo{author}{Chlipala, E.~A.} \emph{et~al.}
\newblock \bibinfo{journal}{\bibinfo{title}{Impact of preanalytical factors during histology processing on section suitability for digital image analysis}}.
\newblock {\emph{\JournalTitle{Toxicologic Pathology}}} \textbf{\bibinfo{volume}{49}}, \bibinfo{pages}{755--772} (\bibinfo{year}{2021}).

\bibitem{chan2014wonderful}
\bibinfo{author}{Chan, J.~K.}
\newblock \bibinfo{journal}{\bibinfo{title}{The wonderful colors of the hematoxylin--eosin stain in diagnostic surgical pathology}}.
\newblock {\emph{\JournalTitle{International journal of surgical pathology}}} \textbf{\bibinfo{volume}{22}}, \bibinfo{pages}{12--32} (\bibinfo{year}{2014}).

\bibitem{bancroft2008theory}
\bibinfo{author}{Bancroft, J.~D.} \& \bibinfo{author}{Gamble, M.}
\newblock \emph{\bibinfo{title}{Theory and practice of histological techniques}} (\bibinfo{publisher}{Elsevier health sciences}, \bibinfo{year}{2008}).

\bibitem{weinstein2009overview}
\bibinfo{author}{Weinstein, R.~S.} \emph{et~al.}
\newblock \bibinfo{journal}{\bibinfo{title}{Overview of telepathology, virtual microscopy, and whole slide imaging: prospects for the future}}.
\newblock {\emph{\JournalTitle{Human pathology}}} \textbf{\bibinfo{volume}{40}}, \bibinfo{pages}{1057--1069} (\bibinfo{year}{2009}).

\bibitem{ji2023physical}
\bibinfo{author}{Ji, X.} \emph{et~al.}
\newblock \bibinfo{journal}{\bibinfo{title}{Physical color calibration of digital pathology scanners for robust artificial intelligence assisted cancer diagnosis}}.
\newblock {\emph{\JournalTitle{arXiv preprint arXiv:2307.05519}}}  (\bibinfo{year}{2023}).

\bibitem{valkonen2020generalized}
\bibinfo{author}{Valkonen, M.}, \bibinfo{author}{H{\"o}gn{\"a}s, G.}, \bibinfo{author}{Bova, G.~S.} \& \bibinfo{author}{Ruusuvuori, P.}
\newblock \bibinfo{journal}{\bibinfo{title}{Generalized fixation invariant nuclei detection through domain adaptation based deep learning}}.
\newblock {\emph{\JournalTitle{IEEE Journal of Biomedical and Health Informatics}}} \textbf{\bibinfo{volume}{25}}, \bibinfo{pages}{1747--1757} (\bibinfo{year}{2020}).

\bibitem{rakha2021current}
\bibinfo{author}{Rakha, E.~A.} \emph{et~al.}
\newblock \bibinfo{journal}{\bibinfo{title}{Current and future applications of artificial intelligence in pathology: a clinical perspective}}.
\newblock {\emph{\JournalTitle{Journal of clinical pathology}}} \textbf{\bibinfo{volume}{74}}, \bibinfo{pages}{409--414} (\bibinfo{year}{2021}).

\bibitem{moscalu2023histopathological}
\bibinfo{author}{Moscalu, M.} \emph{et~al.}
\newblock \bibinfo{journal}{\bibinfo{title}{Histopathological images analysis and predictive modeling implemented in digital pathology—current affairs and perspectives}}.
\newblock {\emph{\JournalTitle{Diagnostics}}} \textbf{\bibinfo{volume}{13}}, \bibinfo{pages}{2379} (\bibinfo{year}{2023}).

\bibitem{wilson2018access}
\bibinfo{author}{Wilson, M.~L.} \emph{et~al.}
\newblock \bibinfo{journal}{\bibinfo{title}{Access to pathology and laboratory medicine services: a crucial gap}}.
\newblock {\emph{\JournalTitle{The Lancet}}} \textbf{\bibinfo{volume}{391}}, \bibinfo{pages}{1927--1938} (\bibinfo{year}{2018}).

\bibitem{strom2020artificial}
\bibinfo{author}{Str{\"o}m, P.} \emph{et~al.}
\newblock \bibinfo{journal}{\bibinfo{title}{Artificial intelligence for diagnosis and grading of prostate cancer in biopsies: a population-based, diagnostic study}}.
\newblock {\emph{\JournalTitle{The Lancet Oncology}}} \textbf{\bibinfo{volume}{21}}, \bibinfo{pages}{222--232} (\bibinfo{year}{2020}).

\bibitem{balkenhol2019deep}
\bibinfo{author}{Balkenhol, M.~C.} \emph{et~al.}
\newblock \bibinfo{journal}{\bibinfo{title}{Deep learning assisted mitotic counting for breast cancer}}.
\newblock {\emph{\JournalTitle{Laboratory investigation}}} \textbf{\bibinfo{volume}{99}}, \bibinfo{pages}{1596--1606} (\bibinfo{year}{2019}).

\bibitem{zadeh2021deep}
\bibinfo{author}{Zadeh~Shirazi, A.} \emph{et~al.}
\newblock \bibinfo{journal}{\bibinfo{title}{A deep convolutional neural network for segmentation of whole-slide pathology images identifies novel tumour cell-perivascular niche interactions that are associated with poor survival in glioblastoma}}.
\newblock {\emph{\JournalTitle{British Journal of Cancer}}} \textbf{\bibinfo{volume}{125}}, \bibinfo{pages}{337--350} (\bibinfo{year}{2021}).

\bibitem{vu2019methods}
\bibinfo{author}{Vu, Q.~D.} \emph{et~al.}
\newblock \bibinfo{journal}{\bibinfo{title}{Methods for segmentation and classification of digital microscopy tissue images}}.
\newblock {\emph{\JournalTitle{Frontiers in bioengineering and biotechnology}}} \bibinfo{pages}{53} (\bibinfo{year}{2019}).

\bibitem{wulczyn2020deep}
\bibinfo{author}{Wulczyn, E.} \emph{et~al.}
\newblock \bibinfo{journal}{\bibinfo{title}{Deep learning-based survival prediction for multiple cancer types using histopathology images}}.
\newblock {\emph{\JournalTitle{PloS one}}} \textbf{\bibinfo{volume}{15}}, \bibinfo{pages}{e0233678} (\bibinfo{year}{2020}).

\bibitem{saillard2020predicting}
\bibinfo{author}{Saillard, C.} \emph{et~al.}
\newblock \bibinfo{journal}{\bibinfo{title}{Predicting survival after hepatocellular carcinoma resection using deep learning on histological slides}}.
\newblock {\emph{\JournalTitle{Hepatology}}} \textbf{\bibinfo{volume}{72}}, \bibinfo{pages}{2000--2013} (\bibinfo{year}{2020}).

\bibitem{fu2020pan}
\bibinfo{author}{Fu, Y.} \emph{et~al.}
\newblock \bibinfo{journal}{\bibinfo{title}{Pan-cancer computational histopathology reveals mutations, tumor composition and prognosis}}.
\newblock {\emph{\JournalTitle{Nature cancer}}} \textbf{\bibinfo{volume}{1}}, \bibinfo{pages}{800--810} (\bibinfo{year}{2020}).

\bibitem{ahmed2022deep}
\bibinfo{author}{Ahmed, A.~A.}, \bibinfo{author}{Abouzid, M.} \& \bibinfo{author}{Kaczmarek, E.}
\newblock \bibinfo{journal}{\bibinfo{title}{Deep learning approaches in histopathology}}.
\newblock {\emph{\JournalTitle{Cancers}}} \textbf{\bibinfo{volume}{14}}, \bibinfo{pages}{5264} (\bibinfo{year}{2022}).

\bibitem{paigeai}
\bibinfo{author}{{Paige AI}}.
\newblock \bibinfo{title}{Paige ai - transforming cancer diagnostics with ai} (\bibinfo{year}{2024}).
\newblock \bibinfo{note}{Accessed: 2024-12-21}.

\bibitem{dunn2024quantitative}
\bibinfo{author}{Dunn, C.} \emph{et~al.}
\newblock \bibinfo{journal}{\bibinfo{title}{Quantitative assessment of h\&e staining for pathology: development and clinical evaluation of a novel system}}.
\newblock {\emph{\JournalTitle{Diagnostic Pathology}}} \textbf{\bibinfo{volume}{19}}, \bibinfo{pages}{42} (\bibinfo{year}{2024}).

\bibitem{breen2024generative}
\bibinfo{author}{Breen, J.}, \bibinfo{author}{Zucker, K.}, \bibinfo{author}{Allen, K.}, \bibinfo{author}{Ravikumar, N.} \& \bibinfo{author}{Orsi, N.~M.}
\newblock \bibinfo{title}{Generative adversarial networks for stain normalisation in histopathology}.
\newblock In \emph{\bibinfo{booktitle}{Applications of Generative AI}}, \bibinfo{pages}{227--247} (\bibinfo{publisher}{Springer}, \bibinfo{year}{2024}).

\bibitem{therrien2018role}
\bibinfo{author}{Therrien, R.} \& \bibinfo{author}{Doyle, S.}
\newblock \bibinfo{title}{Role of training data variability on classifier performance and generalizability}.
\newblock In \emph{\bibinfo{booktitle}{Medical Imaging 2018: Digital Pathology}}, vol. \bibinfo{volume}{10581}, \bibinfo{pages}{58--70} (\bibinfo{organization}{SPIE}, \bibinfo{year}{2018}).

\bibitem{tellez2019quantifying}
\bibinfo{author}{Tellez, D.} \emph{et~al.}
\newblock \bibinfo{journal}{\bibinfo{title}{Quantifying the effects of data augmentation and stain color normalization in convolutional neural networks for computational pathology}}.
\newblock {\emph{\JournalTitle{Medical image analysis}}} \textbf{\bibinfo{volume}{58}}, \bibinfo{pages}{101544} (\bibinfo{year}{2019}).

\bibitem{salvi2023impact}
\bibinfo{author}{Salvi, M.} \emph{et~al.}
\newblock \bibinfo{journal}{\bibinfo{title}{Impact of stain normalization on pathologist assessment of prostate cancer: A comparative study}}.
\newblock {\emph{\JournalTitle{Cancers}}} \textbf{\bibinfo{volume}{15}}, \bibinfo{pages}{1503} (\bibinfo{year}{2023}).

\bibitem{goodfellow2014generative}
\bibinfo{author}{Goodfellow, I.} \emph{et~al.}
\newblock \bibinfo{journal}{\bibinfo{title}{Generative adversarial nets}}.
\newblock {\emph{\JournalTitle{Advances in neural information processing systems}}} \textbf{\bibinfo{volume}{27}} (\bibinfo{year}{2014}).

\bibitem{reinhard2001color}
\bibinfo{author}{Reinhard, E.}, \bibinfo{author}{Adhikhmin, M.}, \bibinfo{author}{Gooch, B.} \& \bibinfo{author}{Shirley, P.}
\newblock \bibinfo{journal}{\bibinfo{title}{Color transfer between images}}.
\newblock {\emph{\JournalTitle{IEEE Computer graphics and applications}}} \textbf{\bibinfo{volume}{21}}, \bibinfo{pages}{34--41} (\bibinfo{year}{2001}).

\bibitem{macenko2009method}
\bibinfo{author}{Macenko, M.} \emph{et~al.}
\newblock \bibinfo{title}{A method for normalizing histology slides for quantitative analysis}.
\newblock In \emph{\bibinfo{booktitle}{2009 IEEE international symposium on biomedical imaging: from nano to macro}}, \bibinfo{pages}{1107--1110} (\bibinfo{organization}{IEEE}, \bibinfo{year}{2009}).

\bibitem{vahadane2016structure}
\bibinfo{author}{Vahadane, A.} \emph{et~al.}
\newblock \bibinfo{journal}{\bibinfo{title}{Structure-preserving color normalization and sparse stain separation for histological images}}.
\newblock {\emph{\JournalTitle{IEEE transactions on medical imaging}}} \textbf{\bibinfo{volume}{35}}, \bibinfo{pages}{1962--1971} (\bibinfo{year}{2016}).

\bibitem{gonzalez2008digital}
\bibinfo{author}{Gonzalez, R.~C.} \& \bibinfo{author}{Woods, R.~E.}
\newblock \bibinfo{journal}{\bibinfo{title}{Digital image processing, prentice hall}}.
\newblock {\emph{\JournalTitle{Upper Saddle River, NJ}}}  (\bibinfo{year}{2008}).

\bibitem{magee2009colour}
\bibinfo{author}{Magee, D.} \emph{et~al.}
\newblock \bibinfo{title}{Colour normalisation in digital histopathology images}.
\newblock In \emph{\bibinfo{booktitle}{Proc Optical Tissue Image analysis in Microscopy, Histopathology and Endoscopy (MICCAI Workshop)}}, vol. \bibinfo{volume}{100}, \bibinfo{pages}{100--111} (\bibinfo{organization}{Daniel Elson London}, \bibinfo{year}{2009}).

\bibitem{niethammer2010appearance}
\bibinfo{author}{Niethammer, M.}, \bibinfo{author}{Borland, D.}, \bibinfo{author}{Marron, J.}, \bibinfo{author}{Woosley, J.} \& \bibinfo{author}{Thomas, N.~E.}
\newblock \bibinfo{title}{Appearance normalization of histology slides}.
\newblock In \emph{\bibinfo{booktitle}{Machine Learning in Medical Imaging: First International Workshop, MLMI 2010, Held in Conjunction with MICCAI 2010, Beijing, China, September 20, 2010. Proceedings 1}}, \bibinfo{pages}{58--66} (\bibinfo{organization}{Springer}, \bibinfo{year}{2010}).

\bibitem{fuchs2011computational}
\bibinfo{author}{Fuchs, T.~J.} \& \bibinfo{author}{Buhmann, J.~M.}
\newblock \bibinfo{journal}{\bibinfo{title}{Computational pathology: challenges and promises for tissue analysis}}.
\newblock {\emph{\JournalTitle{Computerized Medical Imaging and Graphics}}} \textbf{\bibinfo{volume}{35}}, \bibinfo{pages}{515--530} (\bibinfo{year}{2011}).

\bibitem{zhu2017unpaired}
\bibinfo{author}{Zhu, J.-Y.}, \bibinfo{author}{Park, T.}, \bibinfo{author}{Isola, P.} \& \bibinfo{author}{Efros, A.~A.}
\newblock \bibinfo{title}{Unpaired image-to-image translation using cycle-consistent adversarial networks}.
\newblock In \emph{\bibinfo{booktitle}{Proceedings of the IEEE international conference on computer vision}}, \bibinfo{pages}{2223--2232} (\bibinfo{year}{2017}).

\bibitem{levy2020preliminary}
\bibinfo{author}{Levy, J.~J.}, \bibinfo{author}{Jackson, C.~R.}, \bibinfo{author}{Sriharan, A.}, \bibinfo{author}{Christensen, B.~C.} \& \bibinfo{author}{Vaickus, L.~J.}
\newblock \bibinfo{journal}{\bibinfo{title}{Preliminary evaluation of the utility of deep generative histopathology image translation at a mid-sized nci cancer center}}.
\newblock {\emph{\JournalTitle{bioRxiv}}} \bibinfo{pages}{2020--01} (\bibinfo{year}{2020}).

\bibitem{gadermayr2018way}
\bibinfo{author}{Gadermayr, M.}, \bibinfo{author}{Appel, V.}, \bibinfo{author}{Klinkhammer, B.~M.}, \bibinfo{author}{Boor, P.} \& \bibinfo{author}{Merhof, D.}
\newblock \bibinfo{title}{Which way round? a study on the performance of stain-translation for segmenting arbitrarily dyed histological images}.
\newblock In \emph{\bibinfo{booktitle}{Medical Image Computing and Computer Assisted Intervention--MICCAI 2018: 21st International Conference, Granada, Spain, September 16-20, 2018, Proceedings, Part II 11}}, \bibinfo{pages}{165--173} (\bibinfo{organization}{Springer}, \bibinfo{year}{2018}).

\bibitem{koivukoski2023unstained}
\bibinfo{author}{Koivukoski, S.}, \bibinfo{author}{Khan, U.}, \bibinfo{author}{Ruusuvuori, P.} \& \bibinfo{author}{Latonen, L.}
\newblock \bibinfo{journal}{\bibinfo{title}{Unstained tissue imaging and virtual hematoxylin and eosin staining of histologic whole slide images}}.
\newblock {\emph{\JournalTitle{Laboratory Investigation}}} \textbf{\bibinfo{volume}{103}}, \bibinfo{pages}{100070} (\bibinfo{year}{2023}).

\bibitem{khan2023effect}
\bibinfo{author}{Khan, U.}, \bibinfo{author}{Koivukoski, S.}, \bibinfo{author}{Valkonen, M.}, \bibinfo{author}{Latonen, L.} \& \bibinfo{author}{Ruusuvuori, P.}
\newblock \bibinfo{journal}{\bibinfo{title}{The effect of neural network architecture on virtual h\&e staining: Systematic assessment of histological feasibility}}.
\newblock {\emph{\JournalTitle{Patterns}}} \textbf{\bibinfo{volume}{4}} (\bibinfo{year}{2023}).

\bibitem{salido2023comparison}
\bibinfo{author}{Salido, J.}, \bibinfo{author}{Vallez, N.}, \bibinfo{author}{Gonz{\'a}lez-L{\'o}pez, L.}, \bibinfo{author}{Deniz, O.} \& \bibinfo{author}{Bueno, G.}
\newblock \bibinfo{journal}{\bibinfo{title}{Comparison of deep learning models for digital h\&e staining from unpaired label-free multispectral microscopy images}}.
\newblock {\emph{\JournalTitle{Computer Methods and Programs in Biomedicine}}} \textbf{\bibinfo{volume}{235}}, \bibinfo{pages}{107528} (\bibinfo{year}{2023}).

\bibitem{ronneberger2015u}
\bibinfo{author}{Ronneberger, O.}, \bibinfo{author}{Fischer, P.} \& \bibinfo{author}{Brox, T.}
\newblock \bibinfo{title}{U-net: Convolutional networks for biomedical image segmentation}.
\newblock In \emph{\bibinfo{booktitle}{Medical Image Computing and Computer-Assisted Intervention--MICCAI 2015: 18th International Conference, Munich, Germany, October 5-9, 2015, Proceedings, Part III 18}}, \bibinfo{pages}{234--241} (\bibinfo{organization}{Springer}, \bibinfo{year}{2015}).

\bibitem{de2018stain}
\bibinfo{author}{de~Bel, T.}, \bibinfo{author}{Hermsen, M.}, \bibinfo{author}{Kers, J.}, \bibinfo{author}{van~der Laak, J.} \& \bibinfo{author}{Litjens, G.}
\newblock \bibinfo{title}{Stain-transforming cycle-consistent generative adversarial networks for improved segmentation of renal histopathology}.
\newblock In \emph{\bibinfo{booktitle}{International Conference on Medical Imaging with Deep Learning}}, \bibinfo{pages}{151--163} (\bibinfo{organization}{PMLR}, \bibinfo{year}{2019}).

\bibitem{de2021residual}
\bibinfo{author}{de~Bel, T.}, \bibinfo{author}{Bokhorst, J.-M.}, \bibinfo{author}{van~der Laak, J.} \& \bibinfo{author}{Litjens, G.}
\newblock \bibinfo{journal}{\bibinfo{title}{Residual cyclegan for robust domain transformation of histopathological tissue slides}}.
\newblock {\emph{\JournalTitle{Medical Image Analysis}}} \textbf{\bibinfo{volume}{70}}, \bibinfo{pages}{102004} (\bibinfo{year}{2021}).

\bibitem{isola2017image}
\bibinfo{author}{Isola, P.}, \bibinfo{author}{Zhu, J.-Y.}, \bibinfo{author}{Zhou, T.} \& \bibinfo{author}{Efros, A.~A.}
\newblock \bibinfo{title}{Image-to-image translation with conditional adversarial networks}.
\newblock In \emph{\bibinfo{booktitle}{Proceedings of the IEEE conference on computer vision and pattern recognition}}, \bibinfo{pages}{1125--1134} (\bibinfo{year}{2017}).

\bibitem{salehi2020pix2pix}
\bibinfo{author}{Salehi, P.} \& \bibinfo{author}{Chalechale, A.}
\newblock \bibinfo{title}{Pix2pix-based stain-to-stain translation: A solution for robust stain normalization in histopathology images analysis}.
\newblock In \emph{\bibinfo{booktitle}{2020 International conference on machine vision and image processing (MVIP)}}, \bibinfo{pages}{1--7} (\bibinfo{organization}{IEEE}, \bibinfo{year}{2020}).

\bibitem{shaban2019staingan}
\bibinfo{author}{Shaban, M.~T.}, \bibinfo{author}{Baur, C.}, \bibinfo{author}{Navab, N.} \& \bibinfo{author}{Albarqouni, S.}
\newblock \bibinfo{title}{Staingan: Stain style transfer for digital histological images}.
\newblock In \emph{\bibinfo{booktitle}{2019 Ieee 16th international symposium on biomedical imaging (Isbi 2019)}}, \bibinfo{pages}{953--956} (\bibinfo{organization}{IEEE}, \bibinfo{year}{2019}).

\bibitem{cho1710neural}
\bibinfo{author}{Cho, H.}, \bibinfo{author}{Lim, S.}, \bibinfo{author}{Choi, G.} \& \bibinfo{author}{Min, H.}
\newblock \bibinfo{journal}{\bibinfo{title}{Neural stain-style transfer learning using gan for histopathological images. arxiv 2017}}.
\newblock {\emph{\JournalTitle{arXiv preprint arXiv:1710.08543}}}  (\bibinfo{year}{2017]}).

\bibitem{bentaieb2017adversarial}
\bibinfo{author}{BenTaieb, A.} \& \bibinfo{author}{Hamarneh, G.}
\newblock \bibinfo{journal}{\bibinfo{title}{Adversarial stain transfer for histopathology image analysis}}.
\newblock {\emph{\JournalTitle{IEEE transactions on medical imaging}}} \textbf{\bibinfo{volume}{37}}, \bibinfo{pages}{792--802} (\bibinfo{year}{2017}).

\bibitem{cong2021semi}
\bibinfo{author}{Cong, C.} \emph{et~al.}
\newblock \bibinfo{title}{Semi-supervised adversarial learning for stain normalisation in histopathology images}.
\newblock In \emph{\bibinfo{booktitle}{Medical Image Computing and Computer Assisted Intervention--MICCAI 2021: 24th International Conference, Strasbourg, France, September 27--October 1, 2021, Proceedings, Part VIII 24}}, \bibinfo{pages}{581--591} (\bibinfo{organization}{Springer}, \bibinfo{year}{2021}).

\bibitem{he2016deep}
\bibinfo{author}{He, K.}, \bibinfo{author}{Zhang, X.}, \bibinfo{author}{Ren, S.} \& \bibinfo{author}{Sun, J.}
\newblock \bibinfo{title}{Deep residual learning for image recognition}.
\newblock In \emph{\bibinfo{booktitle}{Proceedings of the IEEE conference on computer vision and pattern recognition}}, \bibinfo{pages}{770--778} (\bibinfo{year}{2016}).

\bibitem{international2004image}
\bibinfo{author}{Consortium, I.~C.} \emph{et~al.}
\newblock \bibinfo{journal}{\bibinfo{title}{Image technology colour management-architecture, profile format, and data structure}}.
\newblock {\emph{\JournalTitle{Specification ICC. 1: 2004-10 (Profile version 4.2. 0.0)}}}  (\bibinfo{year}{2004}).

\bibitem{szegedy2016rethinking}
\bibinfo{author}{Szegedy, C.}, \bibinfo{author}{Vanhoucke, V.}, \bibinfo{author}{Ioffe, S.}, \bibinfo{author}{Shlens, J.} \& \bibinfo{author}{Wojna, Z.}
\newblock \bibinfo{title}{Rethinking the inception architecture for computer vision}.
\newblock In \emph{\bibinfo{booktitle}{Proceedings of the IEEE conference on computer vision and pattern recognition}}, \bibinfo{pages}{2818--2826} (\bibinfo{year}{2016}).

\bibitem{heusel2017gans}
\bibinfo{author}{Heusel, M.}, \bibinfo{author}{Ramsauer, H.}, \bibinfo{author}{Unterthiner, T.}, \bibinfo{author}{Nessler, B.} \& \bibinfo{author}{Hochreiter, S.}
\newblock \bibinfo{journal}{\bibinfo{title}{Gans trained by a two time-scale update rule converge to a local nash equilibrium}}.
\newblock {\emph{\JournalTitle{Advances in neural information processing systems}}} \textbf{\bibinfo{volume}{30}} (\bibinfo{year}{2017}).

\bibitem{zhang2011fsim}
\bibinfo{author}{Zhang, L.}, \bibinfo{author}{Zhang, L.}, \bibinfo{author}{Mou, X.} \& \bibinfo{author}{Zhang, D.}
\newblock \bibinfo{journal}{\bibinfo{title}{Fsim: A feature similarity index for image quality assessment}}.
\newblock {\emph{\JournalTitle{IEEE transactions on Image Processing}}} \textbf{\bibinfo{volume}{20}}, \bibinfo{pages}{2378--2386} (\bibinfo{year}{2011}).

\bibitem{prezja2022h}
\bibinfo{author}{Prezja, F.}, \bibinfo{author}{P{\"o}l{\"o}nen, I.}, \bibinfo{author}{{\"A}yr{\"a}m{\"o}, S.}, \bibinfo{author}{Ruusuvuori, P.} \& \bibinfo{author}{Kuopio, T.}
\newblock \bibinfo{journal}{\bibinfo{title}{H\&e multi-laboratory staining variance exploration with machine learning}}.
\newblock {\emph{\JournalTitle{Applied Sciences}}} \textbf{\bibinfo{volume}{12}}, \bibinfo{pages}{7511} (\bibinfo{year}{2022}).

\bibitem{zewdie2021classification}
\bibinfo{author}{Zewdie, E.~T.}, \bibinfo{author}{Tessema, A.~W.} \& \bibinfo{author}{Simegn, G.~L.}
\newblock \bibinfo{journal}{\bibinfo{title}{Classification of breast cancer types, sub-types and grade from histopathological images using deep learning technique}}.
\newblock {\emph{\JournalTitle{Health and Technology}}} \textbf{\bibinfo{volume}{11}}, \bibinfo{pages}{1277--1290} (\bibinfo{year}{2021}).

\bibitem{habtemariam2022cervix}
\bibinfo{author}{Habtemariam, L.~W.}, \bibinfo{author}{Zewde, E.~T.} \& \bibinfo{author}{Simegn, G.~L.}
\newblock \bibinfo{journal}{\bibinfo{title}{Cervix type and cervical cancer classification system using deep learning techniques}}.
\newblock {\emph{\JournalTitle{Medical devices: evidence and research}}} \bibinfo{pages}{163--176} (\bibinfo{year}{2022}).

\bibitem{macqueen1967some}
\bibinfo{author}{MacQueen, J.} \emph{et~al.}
\newblock \bibinfo{title}{Some methods for classification and analysis of multivariate observations}.
\newblock In \emph{\bibinfo{booktitle}{Proceedings of the fifth Berkeley symposium on mathematical statistics and probability}}, \bibinfo{number}{14}, \bibinfo{pages}{281--297} (\bibinfo{organization}{Oakland, CA, USA}, \bibinfo{year}{1967}).

\end{thebibliography}

\end{document}